\documentclass[a4paper,fleqn,usenatbib]{mnras}

\usepackage{amsmath}	
\usepackage{txfonts}

\usepackage[T1]{fontenc}
\usepackage{ae,aecompl}

\usepackage{graphicx,subfig}	
\usepackage{amssymb}	
\usepackage{xcolor}
\usepackage{float}
\newlength{\subcolumnwidth}

\newcommand{\nextsubcolumn}[1][]{%
  \cr\noalign{\hfill}
  \if\relax\detokenize{#1}\relax\else\hsize=#1\setlength{\subcolumnwidth}{\hsize}\fi
}

\title[Investigating star-formation in RCW~42]{Investigating star-formation activity towards the southern HII region RCW~42}

\author[Vipin et al.]{
Vipin Kumar$^{1}$, S. Vig$^{1}$\thanks{Email: sarita@iist.ac.in}, V. S. Veena$^2$, S. Mohan$^1$,
S. K. Ghosh$^{3}$, A. Tej$^{1}$, 
\newauthor
\hspace{0.2cm} and D. K. Ojha$^{3}$ 
\\
$^{1}$Indian Institute of Space Science and Technology, Thiruvananthapuram 695547, India\\
$^{2}$I. Physikalisches Institut, Universit\"at zu K\"oln, Z\"ulpicher Str.77, 50937, K\"oln, Germany\\
$^{3}$Tata Institute of Fundamental Research, Mumbai 400005, India\\
}

\date{Accepted XXX. Received YYY; in original form ZZZ}

\pubyear{2022}
\usepackage{booktabs}
\usepackage{cleveref}
\usepackage{amssymb}
\begin{document}
\label{firstpage}
\pagerange{\pageref{firstpage}--\pageref{lastpage}}
\maketitle

\begin{abstract}
The star-forming activity in the HII region RCW~42 is investigated using multiple wavebands, from near-infrared to radio wavelengths. Located at a distance of 5.8~kpc, this southern region has a bolometric luminosity of 1.8 $\times$ 10$^6$ L$_{\odot}$. The ionized gas emission has been imaged at low radio frequencies of 610 and 1280 MHz using the Giant Metrewave Radio Telescope, India and shows a large expanse of the HII region, spanning $20\times 15$~pc$^2$. The average electron number density in the region is estimated to be $\sim70$~cm$^{-3}$, which suggests an average ionization fraction of the cloud to be $11\%$. An extended green object EGO G274.0649-01.1460 and several young stellar objects have been identified in the region using data from the 2MASS and \textit{Spitzer} surveys. The dust emission from the associated molecular cloud is probed using \textit{Herschel} Space Telescope, which reveals the presence of 5 clumps, C1-C5, in this region.  Two millimetre emission cores of masses $380$ and $390$~M$_\odot$ towards the radio emission peak have been identified towards C1 from the ALMA map at 1.4~mm. The clumps are investigated for their evolutionary stages based on association with various star-formation tracers, and we find that all the clumps are in active/evolved stage.
\end{abstract}




\section{Introduction}\label{introduction}

Massive stars (M~$\gtrsim8$~M$_\odot$) play an important role in the evolution of the interstellar medium due to their high energy output, supernovae explosions, and enrichment of the surrounding medium by heavy elements \citep{2005IAUS..227....3K}. They are believed to form in a manner that is different from their lower mass counterparts \citep{2014prpl.conf..149T,2018ARA&A..56...41M} and are formed in the highest density regions of the molecular clouds. The onset of star-formation in the high density clumps within molecular clouds are heralded by a number of signals, eg.  outflows, masers, hypercompact and ultracompact HII regions, nebulosity associated with embedded clusters. As the massive OB type stars settle into their zero-age main-sequence (ZAMS) phase, an increased output of UV photons takes place which leads to the ionization of the ambient medium and results in the formation of larger HII regions \citep{1999PASP..111.1049G}. The HII regions, therefore, act like radio beacons in molecular clouds to locate massive stars in their infancy.

The star-formation activity in a molecular cloud can be probed by investigating the cloud at multiple wavelengths. This enables the study of various processes responsible for emission at different wavelengths which, in turn, gives important clues about the evolutionary stage of star-formation, i.e. whether the clumps harbour pre-stellar, protostellar or young stellar objects (YSOs). Infrared Dark Clouds are believed to harbour stars in the earliest stages of formation \citep{{2006ApJ...641..389R},{2020ApJ...897...53R}}. As the infant stars evolve, the molecular cloud is affected in multiple ways, eg. by radiation from the stars, the pressure of ionized gas \citep{2015ApJ...815...68M}. We are interested in probing the extent of star formation activity in a molecular cloud that is relatively evolved and hosts an extended HII region. Whether all the clumps in the molecular cloud associated with the HII region, are actively forming stars or whether there exist clumps that are still quiescent is what we are interested in investigating. We have, therefore, considered the Rodgers, Campbell $\&$ Whiteoak (RCW) catalog that compiles nebulous HII regions in the Southern Milky Way which are associated with optical H$\alpha$ emission \citep{1960MNRAS.121..103R}. Although we carry out a multi-wavelength investigation of a single object from this catalog, called RCW42, we believe that this comprehensive study will add and contribute towards understanding the evolution of star formation at the scales of molecular clouds. Our specific goals in the current work are (a) to probe the molecular cloud using dust emission and to estimate the properties (eg. column density, temperature), and (b) study the clumps in the region and the relative evolutionary stages of various clumps using star-formation tracers.

RCW~42, also known as GUM~26 and G274.0-1.1 is a large HII region in the southern sky in the constellation of Vela \citep{{1960MNRAS.121..103R},{2004MNRAS.355..899C}}. The optical images of the emission nebula show high extinction regions criss-crossing the nebula, see Fig.~\ref{radio}(a). The bolometric luminosity of the region is estimated to be 7.2 $\times$ 10$^5$ L$_{\odot}$ by \citet{2014MNRAS.437.1791U}. There is uncertainty regarding distance to this region due to its location ($l=274.01^\circ$), and most kinematic distance estimates in literature range between $\sim 5-6$~kpc \citep{{1998A&AS..132..211H},{2003A&A...397..133R},{2009ApJ...699..469C},{2012ApJ...752..146L},{2014MNRAS.437.1791U}}. In the present work, we adopt a distance of 5.8 kpc which is derived based on the Galactic rotation parameters of \citet{2014ApJ...783..130R} for a $v_{LSR}$ of 37.0 km/s. 

RCW~42 emits strongly in radio and has been detected and studied in multiple Galactic plane surveys and HII region studies \citep{{1972AuJPh..25..443C},{1984MNRAS.210...23G},{1998MNRAS.301..640W},{2011ApJS..195....8C}}. This region hosts two IRAS sources, namely IRAS~09227-5146 and IRAS~09230-5148, and hereafter we refer to them as I-09227 and I-09230, respectively.
High resolution continuum maps ($\sim1.5''$ resolution) of I-09227 at 6.7~GHz obtained by \cite{1998MNRAS.301..640W} reveal two sources separated by $\sim5''$, both with irregular morphologies.
The embedded stellar cluster in this region has been investigated by \cite{2011MNRAS.411..705M} and based on their near-infrared color-magnitude diagrams, they identify few candidate OB stars associated with the near-infrared nebulosity. CO maps presented by \cite{1999A&AS..140..177W} using single-dish measurements provide hints about the associated molecular cloud.

In the present work, we examine this region at radio, infrared and submillimeter wavelengths. We probe the ionized gas emission using low-frequency radio observations from the Giant Metrewave Radio Telescope (GMRT). This enables us to study the compact radio sources as well as the extended diffuse emission. The dust emission from the associated molecular cloud and the clumps are investigated using the far-infrared wavebands. The young stellar objects associated with the cloud have been investigated to identify the exciting sources. 

The organization of this paper is as follows. The details of observations, and archival data are given in Section~\ref{Obs}. The results are presented in Section~\ref{results} followed by a discussion in Section~\ref{discussion}. We finally summarize our conclusions in Section~\ref{conclusion}.

\section{Observations and Data Reduction}\label{Obs}

\subsection{Radio Continuum Observations}\label{Obs_radio}
The radio continuum observations of RCW~42 were carried out at 1280 and 610~MHz using the Giant Metrewave Radio Telescope~\citep{1991CuSc...60...95S}. The GMRT is a fixed array of 30 fully steerable radio antennas, each with a parabolic dish of diameter 45m. Eighteen of these antennas are arranged in a Y-shaped configuration with each arm having six antennas and stretching a length of $\sim14$~km. The remaining twelve antennas are distributed randomly within a central region of area of 1~km$^2$. The minimum and maximum baselines between any two dishes are 100~m and 25~km, respectively, and these decide the range of observable angular scales of target sources. This hybrid configuration enables the imaging of astrophysical objects with high resolution, in addition to sampling diffuse emission at large scales. RCW~42 was observed with GMRT at continuum frequencies of 610 and 1280 MHz with a bandwidth of 32~MHz. The angular extent of the largest structures observable with GMRT are $\sim8\arcmin$ and $\sim20\arcmin$ at 1280 and 610~MHz, respectively. 3C147 was used as the flux calibrator at both the observing frequencies, while 0828-375 and 0837-198 were used as the phase calibrators  at 1280 and 610 MHz, respectively. The details of the observations are listed in Table~\ref{GMRT}.


\begin{table}
\centering
\caption{Details of low frequency radio continuum observations carried out using GMRT.}
\label{GMRT}
\begin{tabular}{lll} 
\toprule
\toprule
Frequency (MHz)      & 610                                        & 1280                                      \\
\toprule
On source time (min) & 191                                        & 174                                       \\
Observation Date     & 2005 November 28                           & 2005 July 14                              \\
Primary Beam (Field-of-view)   &        $44.4'$                   &              $26.2'$        \\
Flux Calibrator      & 3C147                                      & 3C147                                     \\
Phase Calibrator     & 0837-198                                   & 0828-375                                  \\
Synthesized Beam (FWHM)    &            $10''\times 15''$                &     $5''\times9''$           \\
Position Angle       & 0.91$^{\circ}$                                 & -3.42$^{\circ}$                               \\
rms (mJy/beam)     & 1.3                                        & 1.0                                       \\

\bottomrule
\end{tabular}
\end{table}


The data reduction was carried out using the NRAO Astronomical Image Processing System (AIPS). The visibilities corrupted due to non-functional antennas and radio frequency interference were flagged using the tasks {\tt TVFLG} and {\tt UVFLG}. The {\tt SPLIT} command was used to extract out the data of target source which was then deconvolved and self-calibrated using the tasks {\tt IMAGR} and {\tt CALIB}, respectively. RCW~42 is located in the Galactic plane and the large scale emission from the Galactic plane at low radio frequencies contributes as sky noise to the system temperature. As the flux calibrator is located away from the Galactic plane, and observations were carried out with the Automatic Level Corrector (ALC) off, the flux densities of the target source need to be rescaled to account for the change in system temperature due to sky noise from the Galactic plane.  The rescaling factor was calculated using an improved version \citep{2015MNRAS.451.4311R} of the continuum all-sky map of \cite{1982A&AS...47....1H} at 408~MHz. This was used to obtain $T_{Gal}$ at 610 and 1280 MHz using a spectral index of -2.6 for the Galactic plane emission \citep{1999A&AS..137....7R,2011A&A...525A.138G}, and a flux correction corresponding to ${(T_{Gal}\ +\ T_{sys})/T_{sys}}$ was applied to each image. The final images have been treated for primary beam correction using the task {\tt PBCOR}.

\subsection{Far-infrared balloon borne observations} \label{Obs_fir}

The southern Galactic region associated with
RCW~42 was mapped using the 12 channel two band far-infrared (FIR) photometer system (PHT12)
 along with the TIFR 100 cm balloon borne telescope (T100). Details of T100 and PHT12 have  been described elsewhere \citep{{1988ApJ...330..928G},{2000A&A...363..744G}}.
 The observations were carried out during the balloon flight
 launched on 18 November, 1993, from the TIFR Balloon Facility,
 Hyderabad. The region of sky $\sim$ 30$\arcmin$  $\times$ 20$\arcmin$
 around RCW~42 was mapped in the two FIR bands with $\lambda_{eff}$ of 148 \& 209 $\mu$m. The intensity maps in these two FIR bands have been
 generated following the procedure described in \citet{2000A&A...363..744G}. 
 The FIR signals that are sky-chopped were gridded into a matrix with a pixel size of $0.3'\times0.3'$ prior to image processing including deconvolution. The final FIR maps achieve angular resolution of $\sim1'$ and an absolute positional accuracy of $< 0.8'$.


\begin{figure*}
	\includegraphics[height=8cm]{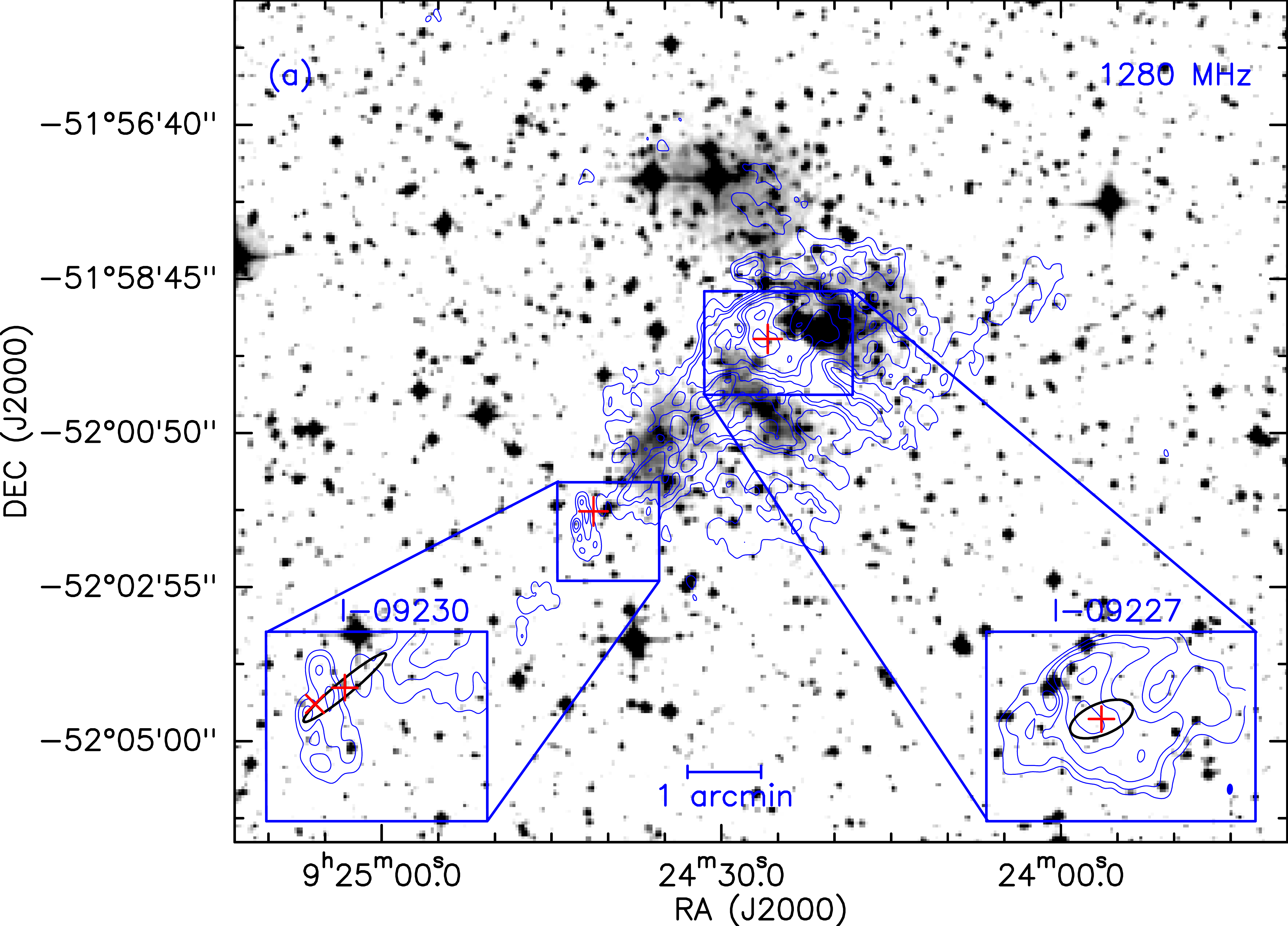}	 
	\hspace*{7mm}
	\includegraphics[height=8cm]{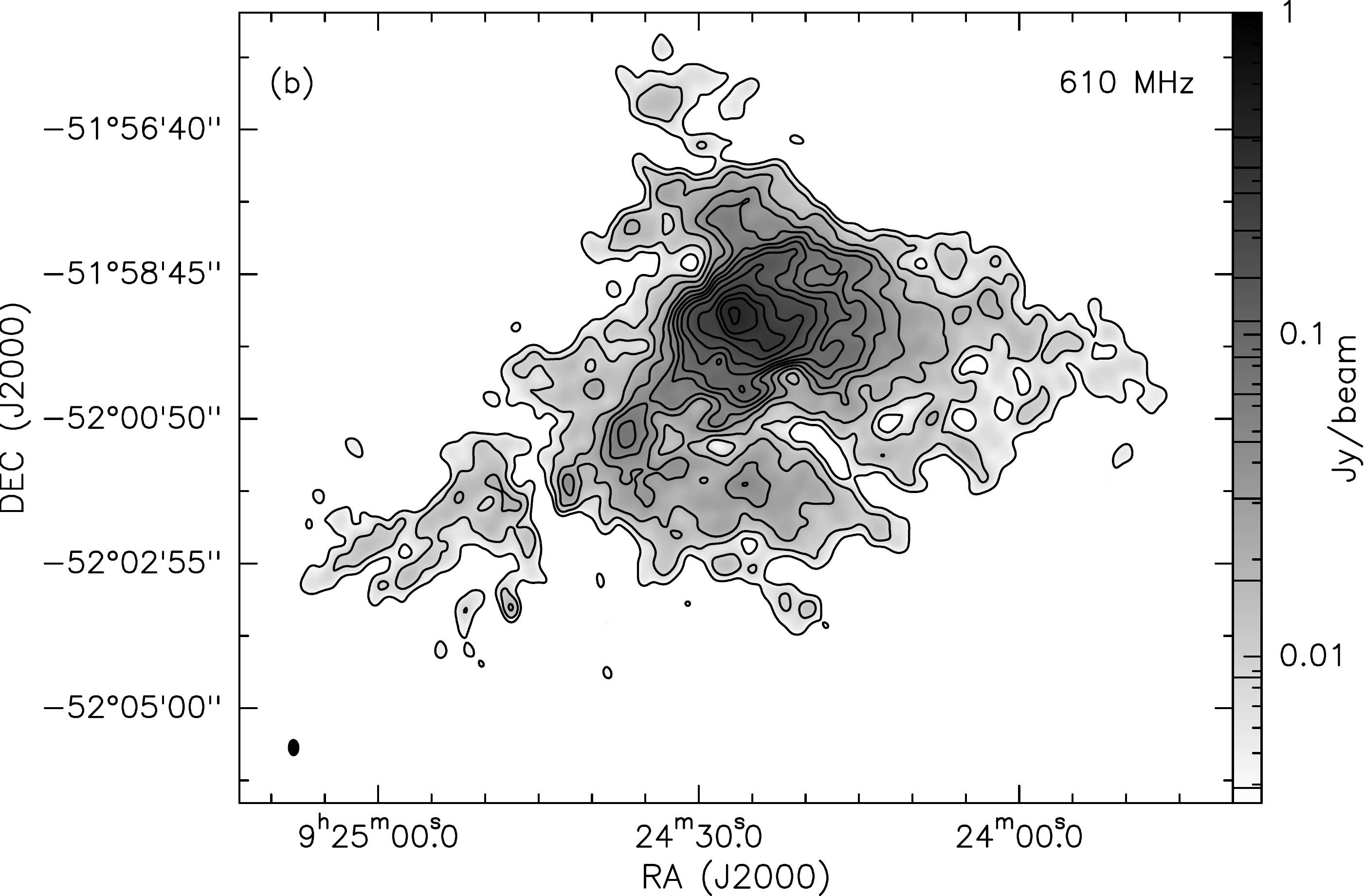}	 
	\caption{Optical and radio continuum emission towards RCW~42. (a) Contours of emission at 1280~MHz overlaid on the DSS optical image, with contour levels at 3.0, 9.2, 15.8, 33.4, 58.7, 85, 140, 215 mJy/beam (beam$\sim5''\times9''$). The positions of the IRAS sources are marked with red + signs and the error ellipses (1$\sigma$) are shown in black. The red $\times$ represents the position of I-09230 from AKARI/IRC survey. (b) Radio continuum emission at 610~MHz (beam$\sim10''\times15''$) and contour levels are marked on the greyscale wedge.}
	\label{radio}
\end{figure*}


\subsection{Archival Data}\label{Obs_archival}

In order to study the HII region in different wavelength bands, archival data from different surveys have been employed. Details about the survey images and catalogs are provided below.

\subsubsection{Two-Micron All Sky Survey (2MASS)}\label{archival_2mass}
The Two Micron All Sky Survey (2MASS) was carried out using two identical 1.3m telescopes located at  Whipple Observatory on Mt. Hopkins, USA, and  Cerro Tololo Inter-American Observatory (CTIO), Chile~\citep{2006AJ....131.1163S}. The survey covered three wavelength bands: J-band (1.235 $\mu$m), H-band (1.662 $\mu$m) and K$\rm{_s}$ band (2.159 $\mu$m). The resolution of the images in all the three bands is $\sim4''$. The Point Source Catalog and images of the 2MASS survey have been used to identify the potential ionizing sources in the molecular cloud.
\subsubsection{Spitzer Space Telescope}\label{archival_spitzer}
In order to probe the warm dust emission and YSOs, images obtained by the InfraRed Array Camera (IRAC) instrument on-board the \textit{Spitzer Space Telescope}~\citep{2004ApJS..154....1W} were used. The IRAC images at 3.6, 4.5, 5.8 and 8.0~$\mu$m were employed for this purpose. The image resolution is $\sim 1.6''$, $1.6''$, $1.8''$ and $1.9''$ at 3.6, 4.5, 5.8 and 8.0~$\mu$m, respectively \citep{2004ApJS..154...10F}. The Vela Carina \citep{meade2014vela} and Deep GLIMPSE \citep{2011sptz.prop80074W} catalogs were used to select point sources in this region, while the images (with pixel size $0.6''$) were extracted from the Spitzer Heritage Archive. 

\subsubsection{Herschel Space Observatory}\label{archival_hershcel}
The cold dust emission from the molecular cloud has been studied using the images from the \textit{Herschel}~Hi-GAL survey \citep{2010PASP..122..314M}. The two instruments used for this survey were the Photometric Array Camera and Spectrometer \cite[PACS;][]{2010A&A...518L...2P}, and the Spectral and Photometric Imaging Receiver \cite[SPIRE;][]{2010A&A...518L...3G} of the \textit{Herschel Space Observatory}. The PACS images are at 70 and 160 $\mu$m, while the Hi-GAL images from SPIRE are at 250, 350 and 500 $\mu$m. We extracted the Level 2.5 images of PACS and SPIRE from Herschel Interactive Processing Environment (HIPE). The beam size (FWHM) of images at 70, 160, 250, 350 and 500~$\mu$m are $5.9^{\prime\prime}$, $11.6^{\prime\prime}$, $18.5^{\prime\prime}$, $25.3^{\prime\prime}$ and $36.9^{\prime\prime}$, respectively. The corresponding pixel sizes are $3.2^{\prime\prime}$, $3.2^{\prime\prime}$, $6^{\prime\prime}$, $10^{\prime\prime}$ and $14^{\prime\prime}$.

\subsubsection{Atacama Large Millimetre/submillimeter Array (ALMA)}\label{archival_alma}
The high resolution cold dust emission towards the peak of the molecular cloud is examined using data from ALMA archive. ALMA is a radio interferometer with antennas of diameter 7m and 12m, located in the Atacama desert, Chile \citep[][]{2003SPIE.4837..110W}. The snapshot observations (120s) of central region of RCW~42 have been carried out as part of the ALMAGAL survey (proposal code 2019.1.00195.6) on Oct~06, 2019, using the 7m array configuration. We used the pipeline reduced continuum image from the archive in the frequency band 216.90 - 220.97~GHz. The pointing observation is towards I-09227 and emission from a region of size (primary beam) $\sim50''$ is measured. The synthesized beam size (FWHM) of the image is $7.5''\times5.0''$, and the rms is 0.8~mJy/beam. We have not considered the associated line emission from spectral cubes as the noise is very high in the snapshot observations.

\section{Results}\label{results}

\subsection{Ionized gas emission}\label{results_ionizedgas}

The radio continuum emission at 1280 and 610 MHz towards RCW~42 is shown in Fig.~\ref{radio}. We see large scale diffuse emission extending in the direction of north-west and south-east, upto $\sim 12'\times9'$, which translates to a size of $\sim 20\times15$~pc$^2$. The large scale morphological shape is similar at both frequencies, with the 610~MHz image being more sensitive to the larger scales as expected from the frequency and antenna configuration. The nebulous radio emission shows large scale extension towards the western side as compared to the eastern side, where the emission gradient is steeper. The higher resolution 1280~MHz image reveals a double peak embedded within a spherical morphology, located towards I-09227. Towards I-09230, we witness radio emission that is elongated north-south. While, the emission lies within the IRAS error ellipse, we note that the coordinates provided by higher resolution AKARI mission for this source \citep{2010A&A...514A...1I} is 13$''$ away and lies within the radio emission. We, therefore, attribute this elongated emission to I-09230. 

The optical emission towards RCW~42 seen in the Digitized Sky Survey image \citep[DSS;][]{1994IAUS..161..167L} is shown in Fig.~\ref{radio}(a). We observe that the radio emission encompasses most of the optical emission of the giant HII region. The high resolution radio image at 1280~MHz exhibits an elongation in the south-west direction that correlates well with that of central extinction ridge in the optical image. Such elongated structures in the radio are often indicative of ionized jets from the central source \citep[e.g.,][]{2016MNRAS.460.1039P}. Diffuse nebulous emission is also seen towards the north-east of I-09227 in the DSS image where we can identify radio emission in the 610~MHz band. There are no IRAS sources associated with this emitting region.

The flux densities of RCW~42 at 1280 and 610~MHz images are $17.9\pm 1.8$~Jy and $25.4\pm2.5$~Jy, respectively, for emission integrated over regions with values above $3\sigma$, where $\sigma$ represents the rms noise. We note that the flux density at 1280 MHz is a lower limit as the largest angular scales that can be mapped at this frequency is $\sim8'$, which is smaller than the size of the region measured at 610 MHz ($\sim12'$). The flux density values are compared with the measurements at other radio frequencies. We have extracted the 843~MHz image of this region from the Sydney University Molonglo Sky Survey \citep[SUMSS;][]{2003MNRAS.342.1117M}, and we obtain a flux density of $18.3\pm1.1$Jy above the $3\sigma$ contour level. \cite{1972AuJPh..25..443C} carried out radio continuum measurements of RCW~42 using the 64m Parkes telescope at 5 and 8.9~GHz and obtained flux densities of 25.9 and 22.4~Jy, respectively.  We note that the flux density measurements are similar at different frequencies (0.6, 5, 9~GHz) suggesting a nearly flat spectral index consistent with a thermally excited HII~region. In order to estimate the percentage of missing flux due to interferometric effects at 1280 MHz, we have considered the fluxes at 3 frequency bands: 610 MHz, 5 GHz and 8.9 GHz that correspond to 25.4, 25.9 and 22.4 Jy, respectively. SUMSS data at 843 MHz is not considered for the analysis as the image is a snapshot and the estimated flux density (18.3 Jy) is therefore a lower limit. Based on the GMRT and Parkes images, we estimate the spectral index of RCW 42 as $-0.03\pm0.04$. Using this spectral index, we estimate the total flux at 1280 MHz as $24.9\pm2.5$~Jy. The percentage of missing flux at 1280 MHz is anticipated to be $\sim$28$\%$.

Using the radio flux density, we next estimate the Lyman continuum photon rate for the entire region using the equation given below \citep{2016A&A...588A.143S}. We, however, note that although this is based on simplistic assumptions such as optically thin emission, negligible absorption of dust, a spherically symmetric HII region and constant density, it provides us with a suggestive estimate of the type of ionising star(s) that could be ionizing the HII region.

\begin{equation}
 {{\dot{N}_{Lyc}}\,\rm{(s^{-1})}} = 4.771\times10^{42} \left(\frac{S_\nu}{\rm Jy} \right)\left(\frac{ T_e}{\rm K} \right)^{-0.45} \left(\frac{\nu}{\rm GHz} \right)^{0.1} \left(\frac{ d}{\rm pc} \right)^{2}
\end{equation}
\noindent Here, $S_\nu$ is the flux density, $T_e$ is the electron temperature, $\nu$ is the frequency of observation and $d$ is the distance to the source.  The electron temperature $T_e$ is taken as 7900~K derived from radio recombination line measurements of this region by \citet{1987A&A...171..261C}. The total flux density of RCW~42 at 1280~MHz, corrected for the missing flux, yields a Lyman continuum photon rate of ${ \dot{N}_{Lyc}}\sim7.2\times10^{49}$~s$^{-1}$. This is equivalent to the Lyman continuum photon rate of a single star of ZAMS type O4 according to \citet{1973AJ.....78..929P}. If we consider the flux density at 5~GHz by \citet{1972AuJPh..25..443C}, we obtain a similar value of ${ \dot{N}_{Lyc}}\sim8.6\times10^{49}$~s$^{-1}$. This corresponds to a single ZAMS star of spectral type O4 \citep{1973AJ.....78..929P} or earlier than O3 \citep{{1996ApJ...460..914V},{2005A&A...436.1049M}}. Earlier studies have indicated the presence of a cluster of OB stars towards this region.  We, therefore, consider a cluster of stars ionizing the gas in the region. The requisite cluster mass can be approximated by using the expression \citep{2019A&A...628A..21W}:
\begin{math}
M_* = 2.19 \times 10^{-47} \,{{\dot{N}_{Lyc}}} 
\end{math}.
\noindent This equation assumes a fully sampled modified Muench initial mass function \citep{2002ApJ...573..366M} and accounts for dust absorption. Using this equation, the total cluster mass is estimated to be $\sim1900$~M$_\odot$ if we consider the flux density at 5~GHz. This would suggest large scale star formation, which is also evident from the size of the HII region.


\begin{figure}
	\includegraphics[height=6.0cm]{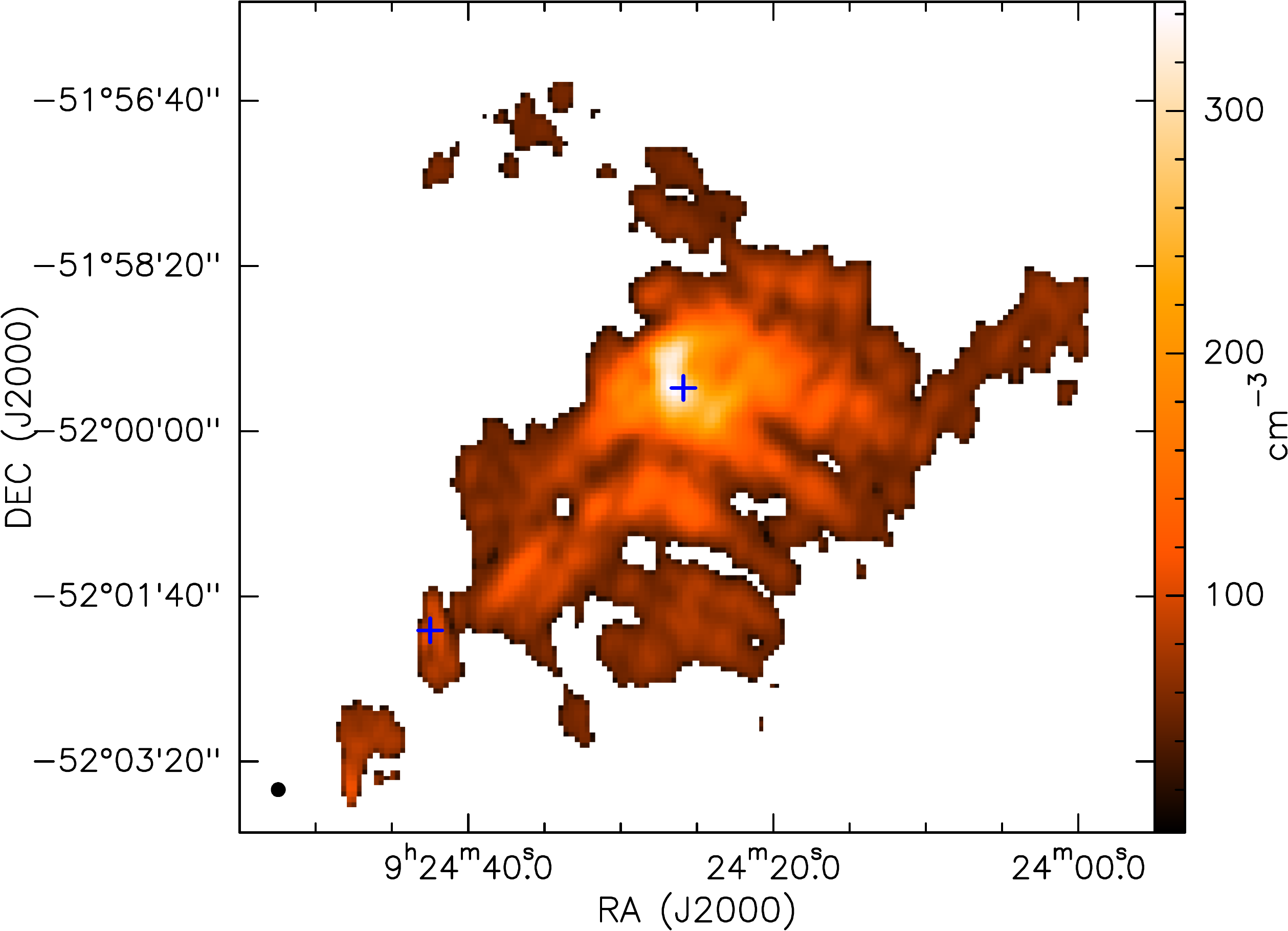}	 
	\caption{Electron density map towards RCW42 constructed using the 1280~MHz radio emission. The positions of I-09227 and I-09230 are marked as blue and cyan + signs, respectively. The map is constructed with beam (FWHM) $\sim9''\times9''$ and pixel size$\sim3''$.}
	\label{ne_radio}
\end{figure}


As RCW~42 spans a large region extending several arcmins, a variation in physical parameters is expected across the region. It is possible to construct a map of $n_e$ for this region by integrating radio flux densities in localised regions across the HII region, and we have utilised the high resolution 1280~MHz map for this purpose. In addition, the 1280 MHz emission has the advantage that the effects of optical depth would be lower compared to 610~MHz. We have carried out a coarse gridding of the RCW42 image into pixels of size $\sim3''\times3''$ that translates to $\sim0.09\times0.09$~pc. We assume that the missing flux corresponding to the large scale emission is uniformly distributed across the region occupied by the 610~MHz emission and this leads to a marginal increase in the flux density of 4.8 mJy per pixel. The average electron density within the regions is estimated from the radio flux density by employing the following expression for thermal bremsstrahlung emission given in cgs units \citep{1986rpa..book.....R}. This expression assumes optically thin emission, and negligible absorption by dust.
\begin{equation}
 { n_e\,\mathrm{(cm^{-3})}} = \left(\frac{1.47\times10^{37}4\pi d^2\, S_\nu \,e^{\mathrm{h}\nu/\mathrm{k_B} T_e}\, {T_e}^{1/2}}{dV\, \bar{g}_{ff}} \right)^{1/2}
\end{equation}
\noindent In this expression, $n_e$ is the electron density, $S_\nu$ represents the flux density from each localised region, $dV$ is the localised volume over which the electron density is determined, ${ \bar{g}_{ff} = \bar{g}_{ff}(\nu,T)}$ is the velocity-averaged Gaunt factor, $\rm{h}$ is the Planck's constant and $\rm k_B$ is the Boltzmann constant. We take ${ \bar{g}_{ff}}(\nu, T_e)\sim5$ as $u = {\mathrm h}\nu/{\mathrm{k_B} T_e} = 8\times10^{-6}$ \citep{1986rpa..book.....R}. We have considered each pixel to be the projection of a localised region whose volume is given by the pixel area multiplied by the line-of-sight distance along the cloud. The line-of-sight distance towards each pixel is calculated assuming that the cloud is spherical of radius $5'$ and centered on I-09227. We have thus created the electron density map which is shown in Fig.~\ref{ne_radio}. The peak electron density of $\sim350$~cm$^{-3}$ is observed towards I-09227, which decreases as we move away from the peak. This would suggest that the ionizing star(s) are within the localised region of I-09227. It is to be noted that these values represent lower limits as the large scale flux is not well sampled at 1280~MHz. The average electron density across the entire region is $\sim70$~cm$^{-3}$.

\subsection{Dust emission}\label{results_dust}
The dust emission towards RCW~42 has been studied using infrared and millimetre continuum images.


\begin{figure}
	\includegraphics[scale=.35]{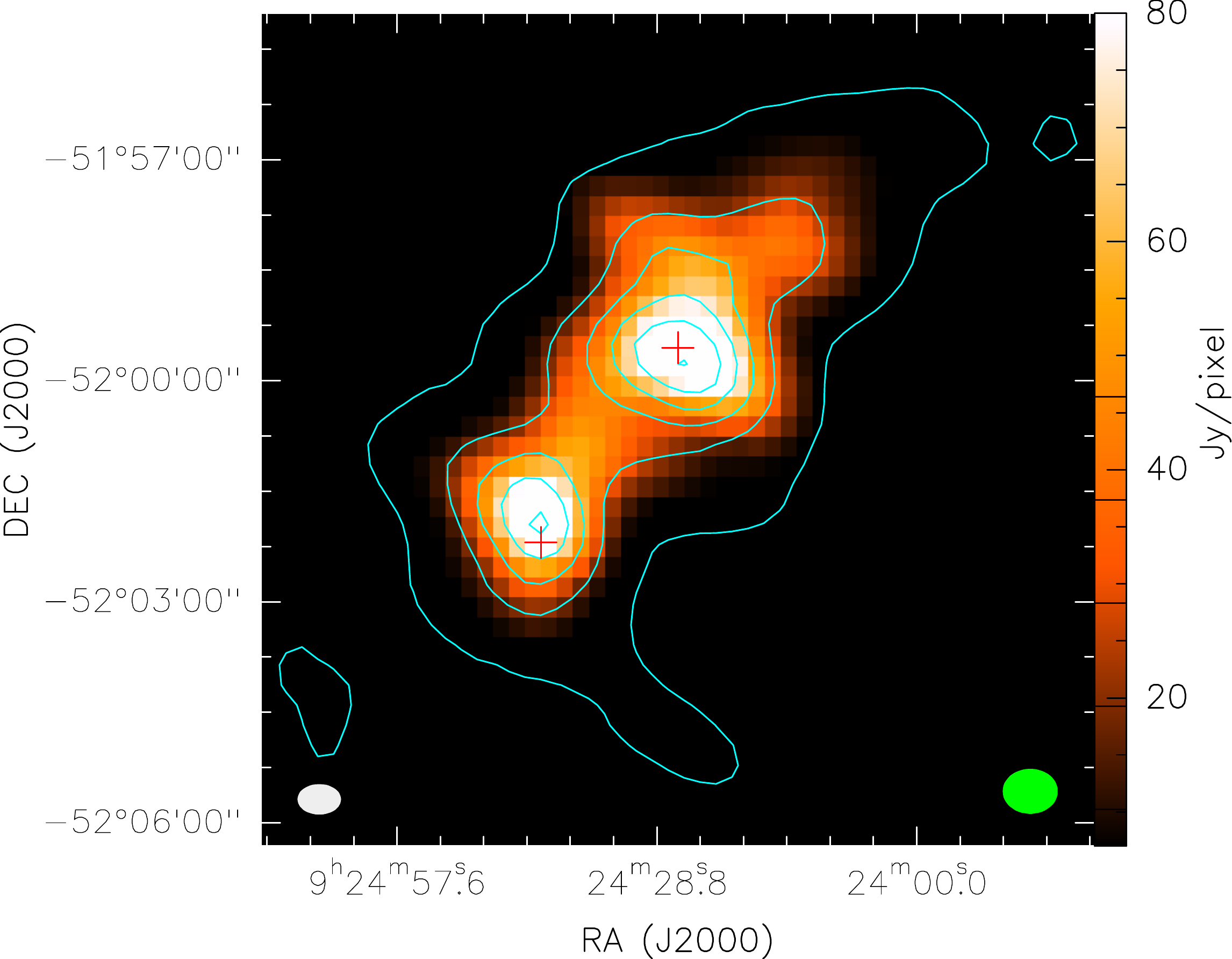}	
	\caption{Far-infrared image of RCW~42 at 148~$\mu$m overlaid with contours of emission at 209~$\mu$m. The images were obtained using the TIFR 100-cm balloon borne telescope. The positions of I-09227 and I-09230 are marked as red + signs. The beamsizes (FWHM) at 148~$\mu$m and 209~$\mu$m are shown by the white and green ellipses, respectively. The contour levels are marked on the wedge.}
\label{t100}
\end{figure}


\subsubsection{Emission from cold dust}
\begin{figure}
	\includegraphics[trim={1 1 1 1},clip,height=7cm]{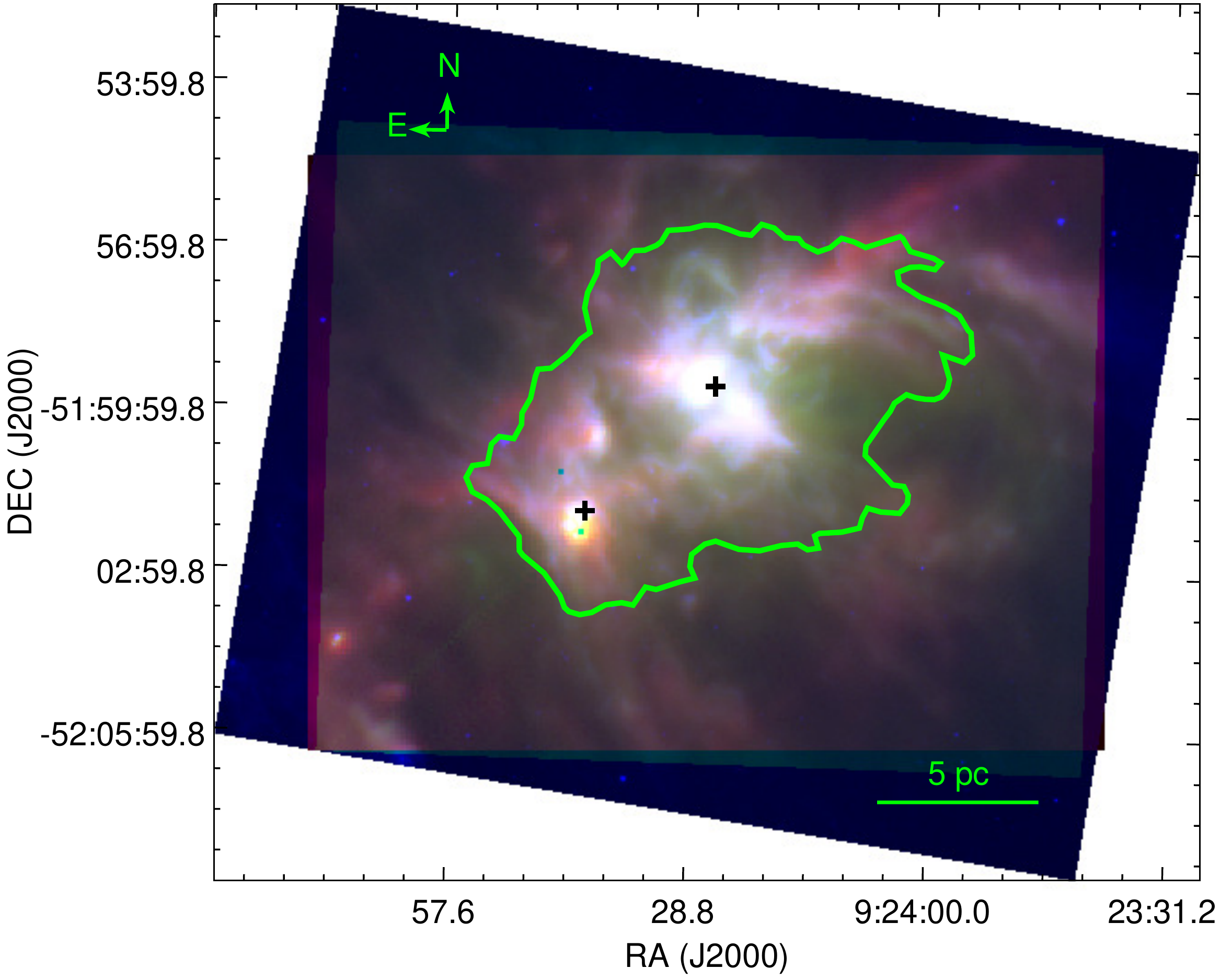}
   \caption{Color composite image of dust emission from RCW~42 including emission at 5.8~$\mu$m (blue), 70~$\mu$m (green) and 250~$\mu$m (red). The contour corresponding to 5.0 Jy/pixel at 70~$\mu$m encloses the region used to estimate the bolometric spectral energy distribution. The positions of I-09227 and I-09230 are marked as + signs.}
   \label{RGB_dust}
\end{figure}

We first describe the emission from the T100 balloon borne images. The images of dust emission at 148 and 209~$\mu$m are shown in Fig.~\ref{t100}. We see emission associated with the sources I-09227 and I-09230. The flux densities at 148 and 209~$\mu$m are determined to be $2500\pm500$~Jy and $1162\pm232$~Jy, respectively. We compare the infrared emission with those from higher resolution Herschel images. The 250~$\mu$m emission, depicted as red in Fig.~\ref{RGB_dust} shows bright compact emission towards the IRAS sources and diffuse emission surrounding this. Faint emission emanating as thin filamentary structures from the entire nebula is also observed.
 
Thermal emission by dust is one of the most efficient methods of estimating gas column densities in molecular clouds, and in many cases even superior to gas tracers as the latter suffer from limitations of excitation conditions \citep{2009ApJ...692...91G}. We therefore, utilise the multiple far-infrared emission maps ($70-500$~$\mu$m) in order to construct images of dust temperature and column density in this region. Studies of star-forming regions have modelled dust emission at 70~$\mu$m as either being due to emission from cold dust \citep{2010A&A...518L..83S}, warm dust \citep{2011A&A...535A..76N}, or having contributions from cold and warm dust \citep{2010A&A...518L..78B}. This is because the emission at 70~$\mu$m could include emission from Very Small Grains \citep[VSGs;][]{2010ApJ...724L..44C}, in addition to emission from large grains. In our study, we have included the 70~$\mu$m emission in the determination of  temperature and column densities as there are regions of active star-formation, eg. location of YSOs. As noted by \citet{2012A&A...542A..10A}, the exclusion of 70~$\mu$m emission is likely to have minimal impact on the determination of dust temperature.


\begin{figure*}
	
	\includegraphics[scale=.3]{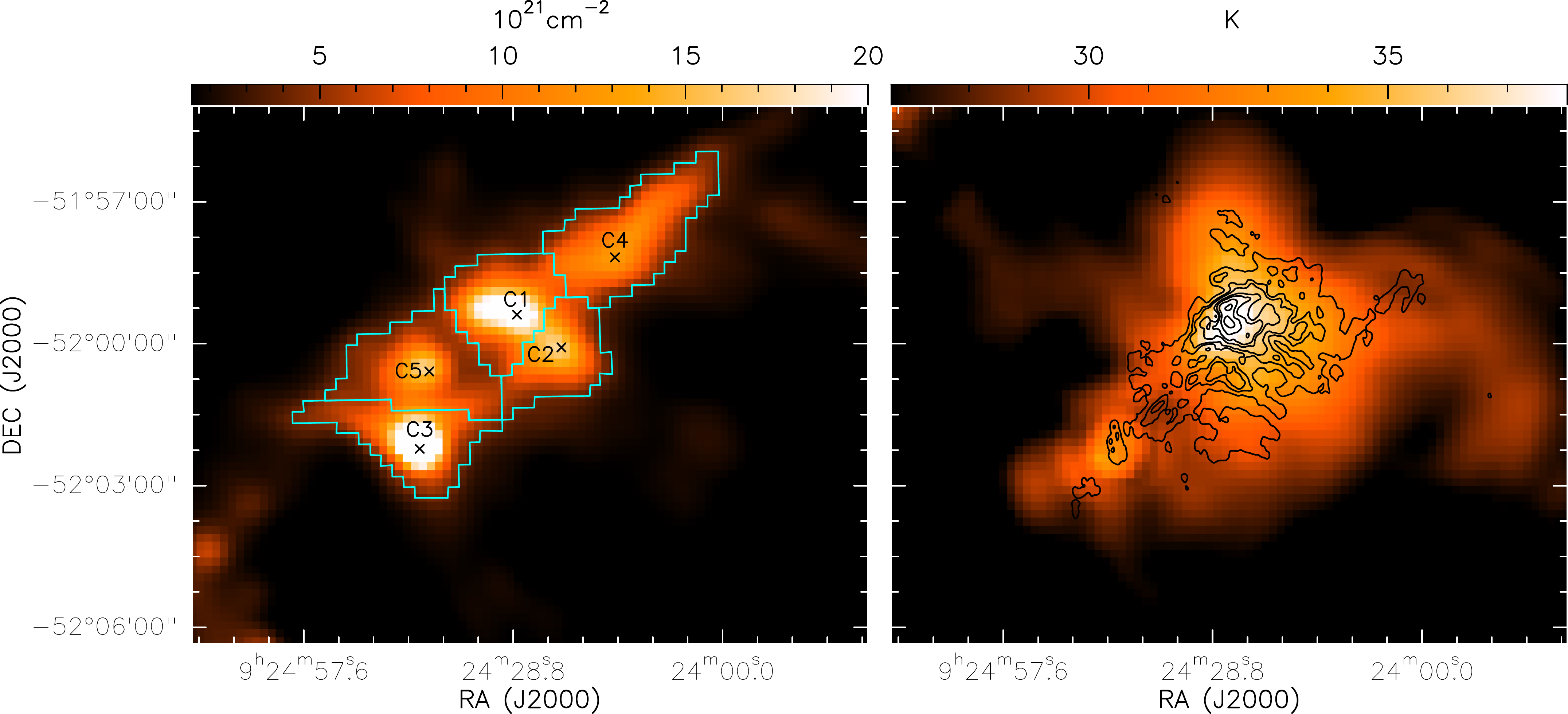}
	\caption{Maps of (a) column density overlaid with clump apertures, and (b) dust temperature overlaid with contours of radio emission at 1280 MHz, the levels are at 3.0, 15.8, 27.8, 58.7, 85, 140, 215 mJy/beam (beam $\sim 5''\times 9''$).}
	\label{dustmaps}
\end{figure*}


We carry out a pixel-to-pixel greybody fitting of the high resolution \textit{Herschel} images and construct line-of-sight averaged molecular hydrogen column density and dust temperature maps. In order to do this, the 70 - 350~$\mu$m images are first convolved and regridded to the lowest resolution (36.9$''$) and pixel size (14$''$) corresponding to the 500~$\mu$m image. The convolution is carried out using kernels from \cite{2011PASP..123.1218A}. For background subtraction of flux densities, we considered multiple nearby regions devoid of emission and determined an average background flux density at each wavelength which is then subtracted from the emission maps. For this analysis, we did not include the T100 flux densities as the images are of lower resolution and sensitivity compared to the \textit{Herschel} images. The spectral energy distributions constructed for every pixel using the background subtracted flux densities, are then fitted to a modified blackbody function of the form \citep{{1990MNRAS.244..458W},{2011A&A...535A.128B},{2013A&A...551A.111O}}:
\begin{equation}
{F_{\nu} = \Omega B_{\nu} (T_{D}) (1-e^{-\tau_{\nu}})}
\label{grey}
\end{equation}
\begin{equation}
{\tau_{\nu} = m_{\rm{H}}\mu N(\rm{H_{2}}) \kappa_{\nu}} 
\label{tau}
\end{equation}
\begin{equation}
{\kappa_{\nu} = 0.1\left(\frac{\nu}{1200\rm{ GHz}}\right)^{\beta}}
\label{opacity}
\end{equation}
Here $F_{\nu}$ is the flux density at frequency $\nu$, $\Omega$ is the solid angle subtended by a pixel, $B_{\nu}$($T_{D}$) is the blackbody function at dust temperature $T_D$, $N(\rm{H_2})$ is the column density of molecular hydrogen, and $m_{\rm{H}}\mu$ is the mean particle mass with $m_{\rm{H}}$ being the mass of a hydrogen atom. $\mu$ is taken to be 2.86 assuming a gas composition of 70\% molecular hydrogen by mass \citep{2010A&A...518L..92W}. The opacity per cloud mass is denoted by $\kappa_\nu$ which assumes a certain gas-to-dust ratio by mass, and $\beta$ is the dust emissivity index. We have considered $\kappa_\nu\sim0.1$~cm$^2$/g at 250~$\mu$m, uncertain upto a factor of 2 \citep{1983QJRAS..24..267H} for a gas-to-dust mass ratio of 100. The best fits were obtained using non-linear least-squares Levenberg-Marquardt algorithm. We considered the dust emissivity index, $\beta$ equal to 2.0 and assumed flux uncertainties of $\sim15\%$ in the bands \citep{{2009A&A...504..415S},{2013A&A...551A..98L}}. 

The column density and dust temperature maps are shown in Fig.~\ref{dustmaps}. The hierarchical structure of the cloud is evident from the contours of column density where clumpy structures are located within the larger molecular cloud \citep{1992ApJ...393..172H}. The highest column density pixel is observed towards I-09230 with a value of $5.8\pm0.6\times10^{22}$~cm$^{-2}$ corresponding to a temperature of $32.9\pm1.0$~K. The dust temperature map shows an extended distribution that correlates well with the ionized gas emission. The peak temperature of $39.1\pm1.3$~K is observed towards I-09227, with a column density of $2.0\pm0.2\times10^{22}$~cm$^{-2}$. We next study the small scale structures within the cloud, denoted as clumps.

\begin{table*}
\caption{Physical parameters of the five clumps identified towards RCW~42.}
{\setlength{\tabcolsep}{0.1mm}
\begin{tabular}{cccccccccc}
\toprule
\toprule
Clump & \begin{tabular}[c]{@{}c@{}}$\alpha_\textrm{J2000}$\\ ($^{h\,m\,s}$)\end{tabular} & \begin{tabular}[c]{@{}c@{}}$\delta_\textrm{J2000}$\\ ($^{\circ\,'\,''}$)\end{tabular} & \begin{tabular}[c]{@{}c@{}}Area\\ (pc$^2$)\end{tabular} & \begin{tabular}[c]{@{}c@{}}Average\\ (Peak)\\ Temperature\\ (K)\end{tabular} & \begin{tabular}[c]{@{}c@{}}Average\\ (Peak)\\ Column\\ density\\ ($\times10^{21}$\\ cm $^{-2}$)\end{tabular} & \begin{tabular}[c]{@{}c@{}}Total\\ Gas\\ Mass\\ ($\times10^2$\\ M$_\odot$)\end{tabular} & \begin{tabular}[c]{@{}c@{}}FIR \\ Luminosity\\ ($\times10^3$ L$_\odot$)\end{tabular} & \begin{tabular}[c]{@{}c@{}}Ionized\\ Gas\\ Mass\\ ($\times10^2$\\ M$_\odot$)\end{tabular} & \begin{tabular}[c]{@{}l@{}}Ionization\\ Fraction\end{tabular} \\
\midrule
C1    & \begin{tabular}[c]{@{}c@{}}09:24:28.28\end{tabular}                                                                      & \begin{tabular}[c]{@{}c@{}}$-51$:59:23.67\end{tabular}                                                                        & \begin{tabular}[c]{@{}c@{}}12.4\end{tabular}                                                     & \begin{tabular}[c]{@{}c@{}}34.2 $\pm$ 1.0 (39.1)\end{tabular}                                         & \begin{tabular}[c]{@{}c@{}}12.0 $\pm$ 1.2 (26.1)\end{tabular}                                                                                    & \begin{tabular}[c]{@{}c@{}}34\end{tabular}                                                                                                                                                                        & \begin{tabular}[c]{@{}c@{}}295\end{tabular}                                                                                    & \begin{tabular}[c]{@{}c@{}}4.11\end{tabular}                                                                                   & \begin{tabular}[c]{@{}c@{}}0.11\end{tabular}                                                          \\
C2    & \begin{tabular}[c]{@{}c@{}}09:24:22.17\end{tabular}                                                                      & \begin{tabular}[c]{@{}c@{}}$-52$:00:05.27\end{tabular}                                                                        & \begin{tabular}[c]{@{}c@{}}9.9\end{tabular}                                                     & \begin{tabular}[c]{@{}c@{}}34.4 $\pm$ 1.0 (37.7)\end{tabular}                                                      &\begin{tabular}[c]{@{}c@{}}9.6 $\pm$ 0.9 (17.4)\end{tabular}                                                                                    & \begin{tabular}[c]{@{}c@{}}22\end{tabular}                                                                                     & \begin{tabular}[c]{@{}c@{}}197\end{tabular}                                                                     & \begin{tabular}[c]{@{}c@{}}4.46\end{tabular}                                                                                   & \begin{tabular}[c]{@{}c@{}}0.19\end{tabular}                                                          \\
C3    & \begin{tabular}[c]{@{}c@{}}09:24:41.66\end{tabular}                                                                      & \begin{tabular}[c]{@{}c@{}}$-52$:02:13.91\end{tabular}                                                                        & \begin{tabular}[c]{@{}c@{}}12.6\end{tabular}                                                     & \begin{tabular}[c]{@{}c@{}}30.1 $\pm$ 0.7 (34.1)\end{tabular}                                                      &\begin{tabular}[c]{@{}c@{}}14.0 $\pm$ 1.1 (58.3)\end{tabular}                                                                                    & \begin{tabular}[c]{@{}c@{}}40\end{tabular}                                                                                     & \begin{tabular}[c]{@{}c@{}}177\end{tabular}                                                                                    & \begin{tabular}[c]{@{}c@{}}0.92\end{tabular}                                                                                   & \begin{tabular}[c]{@{}c@{}}0.02\end{tabular}                                                          \\
C4    & \begin{tabular}[c]{@{}c@{}}09:24:14.82\end{tabular}                                                                      & \begin{tabular}[c]{@{}c@{}}$-51$:58:10.89\end{tabular}                                                                        & \begin{tabular}[c]{@{}c@{}}17.7\end{tabular}                                                     & \begin{tabular}[c]{@{}c@{}}28.9 $\pm$ 0.9 (34.2)\end{tabular}                                                      &\begin{tabular}[c]{@{}c@{}}8.0 $\pm$ 0.9 (13.0)\end{tabular}                                                                                    & \begin{tabular}[c]{@{}c@{}}32\end{tabular}                                                                                     & \begin{tabular}[c]{@{}c@{}}113\end{tabular}                                                                                     & \begin{tabular}[c]{@{}c@{}}1.23\end{tabular}                                                                                   & \begin{tabular}[c]{@{}c@{}}0.03\end{tabular}                                                          \\
C5    & \begin{tabular}[c]{@{}c@{}}09:24:40.34\end{tabular}                                                                      & \begin{tabular}[c]{@{}c@{}}$-52$:00:35.70\end{tabular}                                                                        & \begin{tabular}[c]{@{}c@{}}16.0\end{tabular}                                                     & \begin{tabular}[c]{@{}c@{}}28.3 $\pm$ 0.8 (30.4)\end{tabular}                                                      &\begin{tabular}[c]{@{}c@{}}8.0 $\pm$ 0.9 (17.0)\end{tabular}                                                                                    &\begin{tabular}[c]{@{}c@{}}29\end{tabular}                                                                                     & \begin{tabular}[c]{@{}c@{}}92\end{tabular}                                                                                     & \begin{tabular}[c]{@{}c@{}}1.47\end{tabular}                                                                                   & \begin{tabular}[c]{@{}c@{}}0.05\end{tabular}                                                         
\\ \bottomrule                                                                           
\end{tabular}}
\label{clumpcharac}
\end{table*}

\subsubsection{Properties of Clumps}\label{dust_clumps}

Clumps represent inherent high density and bound structures within molecular clouds that are potential sites of star cluster formation \citep{2000prpl.conf...97W}. Clumps are generally classified as structures on the size scale of parsecs, mass of few hundred solar masses and densities of the order of 10$^4$~cm$^{-3}$ \citep{2018ARA&A..56...41M}. 
To ascertain the properties of higher density structures in the cloud, we use the algorithm Clumpfind \citep{1994ApJ...428..693W} on the column density map. The algorithm employs contours for clump identification and therefore suitable threshold and spacing levels are essential for identification of all the clumps in the region. In the present case, we select a threshold level of $5\sigma$ that corresponds to $N(\mathrm{H_2}) = 4.1\times10^{21}$~cm$^{-2} $,  and a contour spacing of $3\sigma$ for the detection of clumps, where $\sigma = 8.2\times10^{20}$~cm$^{-2}$. 

We distinguish five clumps in the region. These are designated as  C1 - C5 and shown in the column density map in Fig.~\ref{dustmaps}(a). The peak and average column densities and dust temperature values are listed in Table~\ref{clumpcharac}.  The values of average column density and dust temperature in the clumps are found to be similar: $N$(H$_2$)$\sim0.8-1.40\times10^{22}$~cm$^{-2}$, and $T_D\sim28-34$~K. Among the clumps, C1 and C2 have the highest average temperature, while C3 displays the largest average column density. The IRAS source I-09227 is located within Clump C1 while I-09230 is located in C3. 

In addition to column density and temperature, we also determine the mass and luminosity of the clumps. The estimated masses of clumps correspond to the total gas mass in the clumps which will be compared with the ionized gas mass later. The total gas mass, $M_c$, in each clump is estimated using the following equation.
\begin{equation}
{M_c = N(\mathrm{H_2})\mu m_{\mathrm{H}} A}
\label{cmass}
\end{equation}
Here, $A$ is its physical area determined by the aperture. The clumps have mass in the range ${2200-4000}$~M$_\odot$, and the total mass of the clumps is $1.6\times10^4$~M$_\odot$. These are also listed in Table~\ref{clumpcharac}. As mentioned earlier, the values estimated in this way are uncertain to a factor of 2, and therefore, the column densities and masses of clumps derived are uncertain upto $\sim50$\% of their values. The uncertainity in total gas mass estimation mainly arises due to our lack of understanding regarding accurate values of $\kappa_\nu$ and $\beta$ at the frequencies considered. We have, therefore, also considered a cloud opacity of $6.8\times10^{-3}$~cm$^2$g$^{-1}$ at 1.2~mm considering a gas-to-dust mass ratio of $\sim150$ \citep{2013IAUS..292...19T} and a value of dust emissivity index $\beta=1.8$ determined from Galactic plane emission \citep{2011A&A...536A..25P}. We carry out the pixel-to-pixel SED fitting with these values anew. We estimate the column densities and masses, and find that the overall clump densities and masses are lower by about ${25-30\%}$ compared to the former case, and total mass of clumps is ${1.2\times10^4}$~M$_\odot$. This is within the uncertainty range given above.

The luminosity of each clump is estimated by integrating under the best-fit greybody to the SED (constructed using flux densities within clump apertures) using the method described earlier for each pixel. The luminosities of the clumps lie in the range ${90-300\times10^3}$~L$_\odot$, and the total luminosities of the clumps is ${8.7\times10^5}$~L$_\odot$. This represents the far-infrared luminosity of the high density clumps, and a better assessment of the bolometric luminosity of the region can be obtained by considering the entire region sampling the diffuse cloud with a broader wavelength coverage. This is discussed in Sect. 3.3. 

\subsubsection{Millimetre continuum emission towards C1}

\begin{figure}
    \hskip -0.5cm
    \includegraphics[scale=0.38]{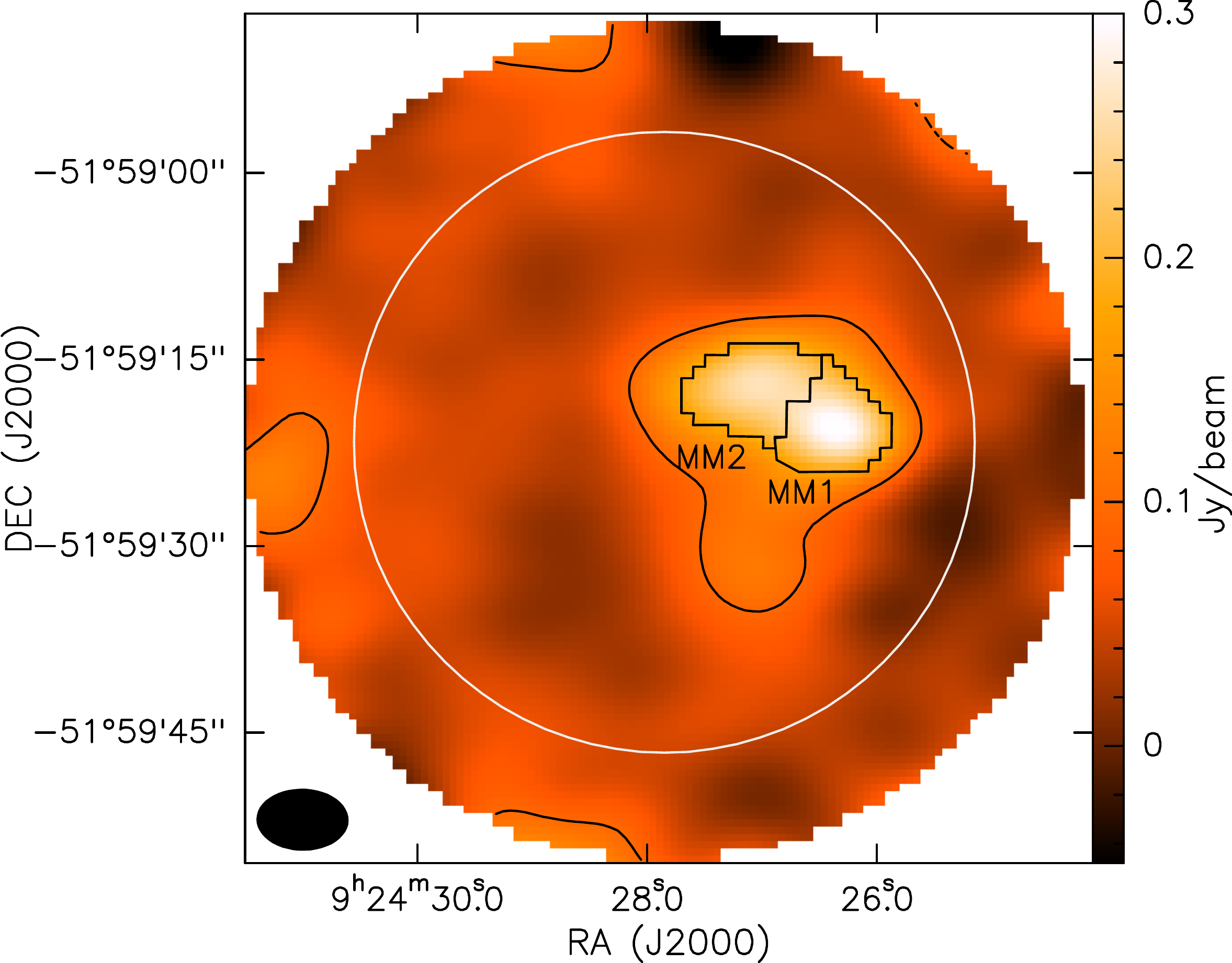}
    \caption{1.4~mm continuum map of RCW~42 towards I-09227 obtained from ALMA. Apertures of the two cores MM1 and MM2 are shown. The primary beam (field-of-view) is marked with a white circle, and the synthesized beam (FWHM) is shown towards the bottom left. The contour corresponds to an emission of 61.0 mJy/beam (beam $\sim 5''\times 9''$).}
\label{cont1}
\end{figure}

The ALMA 1.4~mm continuum image covers the region around the clump C1. We have identified two cores within clump C1 (Fig.~\ref{cont1}), using the algorithm Clumpfind. A threshold of 0.15~Jy/beam and a contour spacing of 7~mJy/beam have been used to identify the dusty cores. The two cores, shown in Fig.~\ref{cont1}, are labelled MM1 and MM2, the former being more luminous than the latter. The two cores are separated by $7.7''\sim0.2$ pc. We have estimated the masses of the cores assuming that the emission is optically thin at this frequency. The masses are quite similar: 380 and 390~M$_\odot$ for MM1 and MM2, respectively. To obtain these estimates, we have considered the gas-to-dust ratio of 100, and opacity per cloud mass of 0.3~cm$^2$/g, using the relation given by \citet{1983QJRAS..24..267H} for $\beta=2$. The temperature of 38~K is adopted from the dust temperature map. Using the opacity relation from \citet{2013IAUS..292...19T} for $\beta=1.8$ with gas-to-dust ratio of 150, we estimate the dust masses to be 250 and 260~M$_\odot$, respectively. From the continuum emission, we observe an extension towards the south of MM1 ($\sim 13''$) and it is possible that this represents a third core. With a higher signal-to-noise ratio image, it would be possible to confirm this. 


\subsubsection{Emission from warm dust}

\begin{figure}
	\includegraphics[trim={1 1 1 1},clip,height=6.5cm]{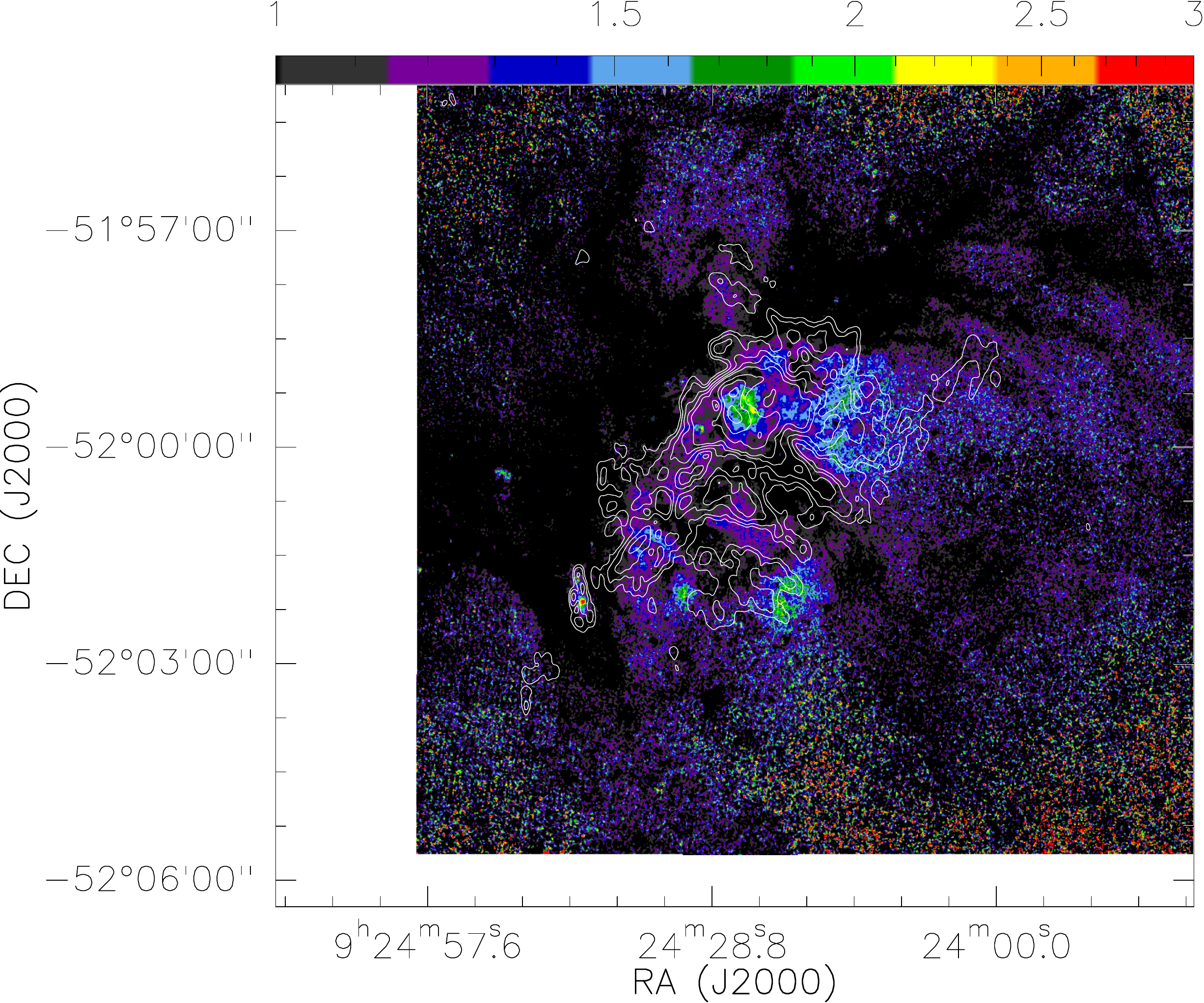}
	\caption{Flux ratio map of \textit{Spitzer}[4.5]/[3.6] overlaid with 1280~MHz radio contours. The contour levels are 3.0, 7.2, 15.8, 27.9, 40.8, 58.7, 85, 140, 215 mJy/beam.}
	\label{irac_flux_ratio}
\end{figure}

The warm dust emission from this region at 5.8~$\mu$m and 70~$\mu$m is shown in Fig.~\ref{RGB_dust}. The emission at mid-infrared wavelengths can be primarily attributed to stochastically heated small dust grains \citep{2003ARA&A..41..241D} that absorb UV radiation from the ionized region as well as thermal emission from warm dust in the circumstellar envelope. Additionally, the 3.6, 5.8 and 8.0~$\mu$m bands contain emission features from polycyclic aromatic hydrocarbons (PAH), that are excited by UV photons in the photodissociation regions (PDR). The 4.5~$\mu$m band, on the other hand, is dominated by shocked H$_2$ and CO emission \citep{{2004ApJS..154..352N},{2008MNRAS.384...71T}} that are associated with protostellar outflows. A mid-infrared flux ratio map constructed using the 3.6~$\mu$m and the 4.5~$\mu$m images can be utilised to discern regions where outflows are present. This is possible because the [4.5]$/$[3.6] ratio is large ($>1.5$) for jets and outflows whereas it is lower for regions dominated by stellar sources \citep{2010ApJ...720..155T}. The [4.5]$/$[3.6] flux ratio image of the RCW 42 region is shown in Fig.~\ref{irac_flux_ratio}. The ratio image shows striking features with an excess in [4.5]/[3.6] of values larger than 1.5 towards the central region, close to the radio peak as well as the outer regions associated with the diffuse radio emission. The flux ratio enhancement towards the main radio peak and localized excess towards other radio peaks could be indicative of outflow activity from the embedded massive young stellar objects, whereas the more diffuse 4.5~$\mu$m excess towards the outer regions, i.e., west and south-west, could originate from shocks resulting from the expansion of the HII~region. An enhancement in flux ratio ($1<[4.5/[3.6]<1.5]$) is also observed towards the nebulous emission of I-09227, seen in the optical image. Though the flux ratio in this region is lower than that expected for shocks, the enhancement along with the presence of radio and optical emission suggests the vigorous star formation activity in the region.


\begin{figure}
	\includegraphics[trim={1 1 1 1},clip,height=7.5cm]{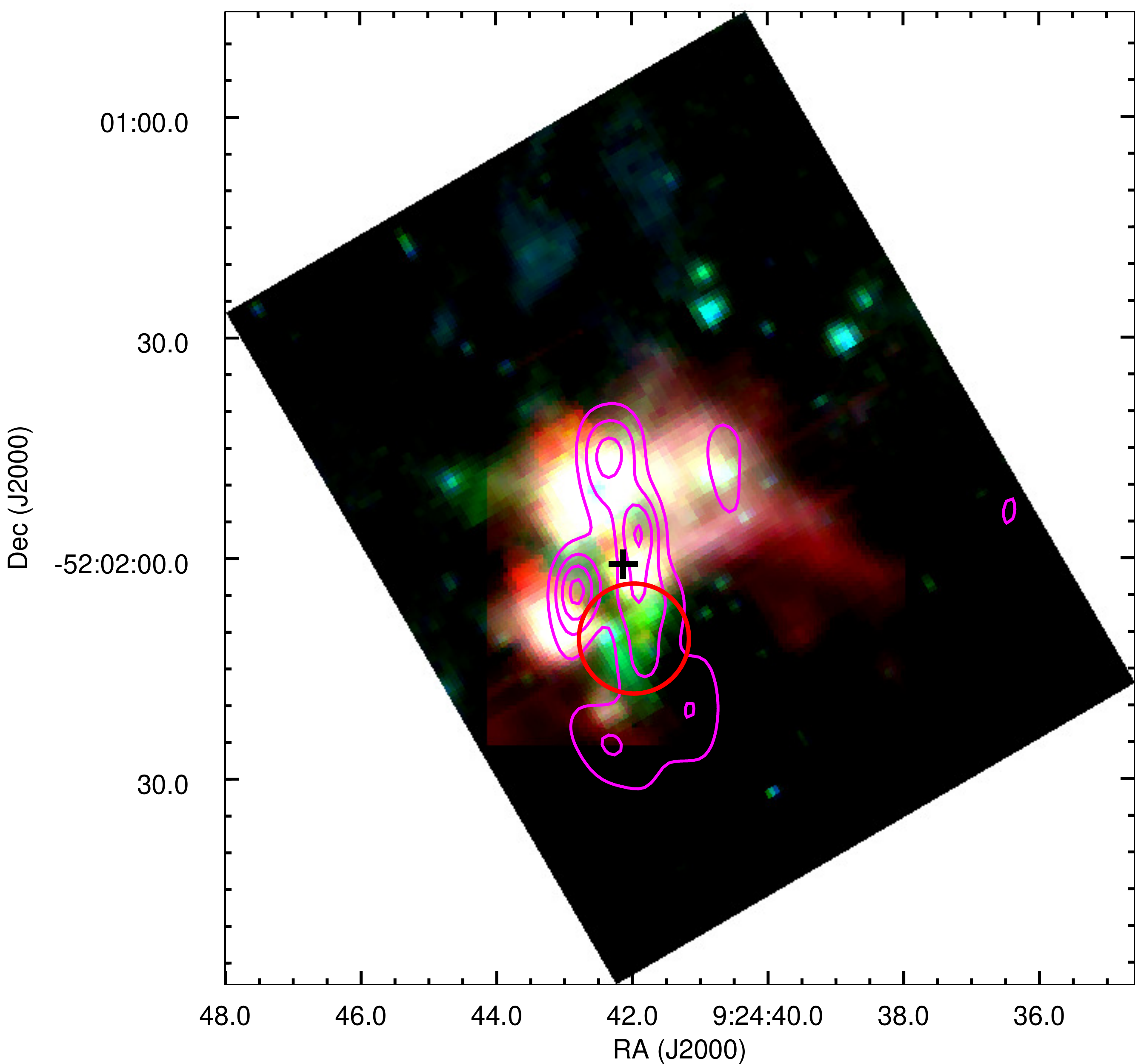}
	\caption{Mid-infrared color composite map (red: 8~$\mu$m, green: 4.5~$\mu$m, blue: 3.6~$\mu$m) of I-09230 overlaid with 1280 MHz radio contours. The contour levels are 5, 10, 15, 20 mJy/beam. The location of the massive YSO outflow candidate G274.0649--01.1460 is marked as a + sign and the associated green fuzzies are enclosed within a circle (cyan).}
	\label{EGO_1280}
\end{figure}

A visual inspection of the \textit{Spitzer} three-color composite map (Fig.~\ref{EGO_1280}) shows an enhancement in 4.5~$\mu$m emission towards I-09230 where there are 2 green fuzzies associated with the mid-infrared point sources. Such objects identified as enhanced green emission in \textit{Spitzer} RGB maps are classified as extended green objects  \citep[EGO;][]{2008AJ....136.2391C}. They are believed to trace the shocked emission from massive protostellar outflows. This EGO is also associated with radio emission with an elongated morphology, indicative of ionized jet, evident from Fig.~\ref{radio}(b). The [4.5]/[3.6] ratio along the elongated radio emission is high ($>$3), characteristic of outflows. The mid-infrared point source associated with the radio emission, G274.0649--01.1460 is classified as a candidate YSO by \citet{2007A&A...476.1019M} and as a high mass YSO outflow candidate \citep[][and references therein]{2012ApJ...753...51G}. We refer to this source as EGO G274.0649--01.1460 in this work.

\subsection{Bolometric Luminosity of RCW~42}\label{bol_lum}

\begin{figure}
\begin{center}
\includegraphics[width=8cm]{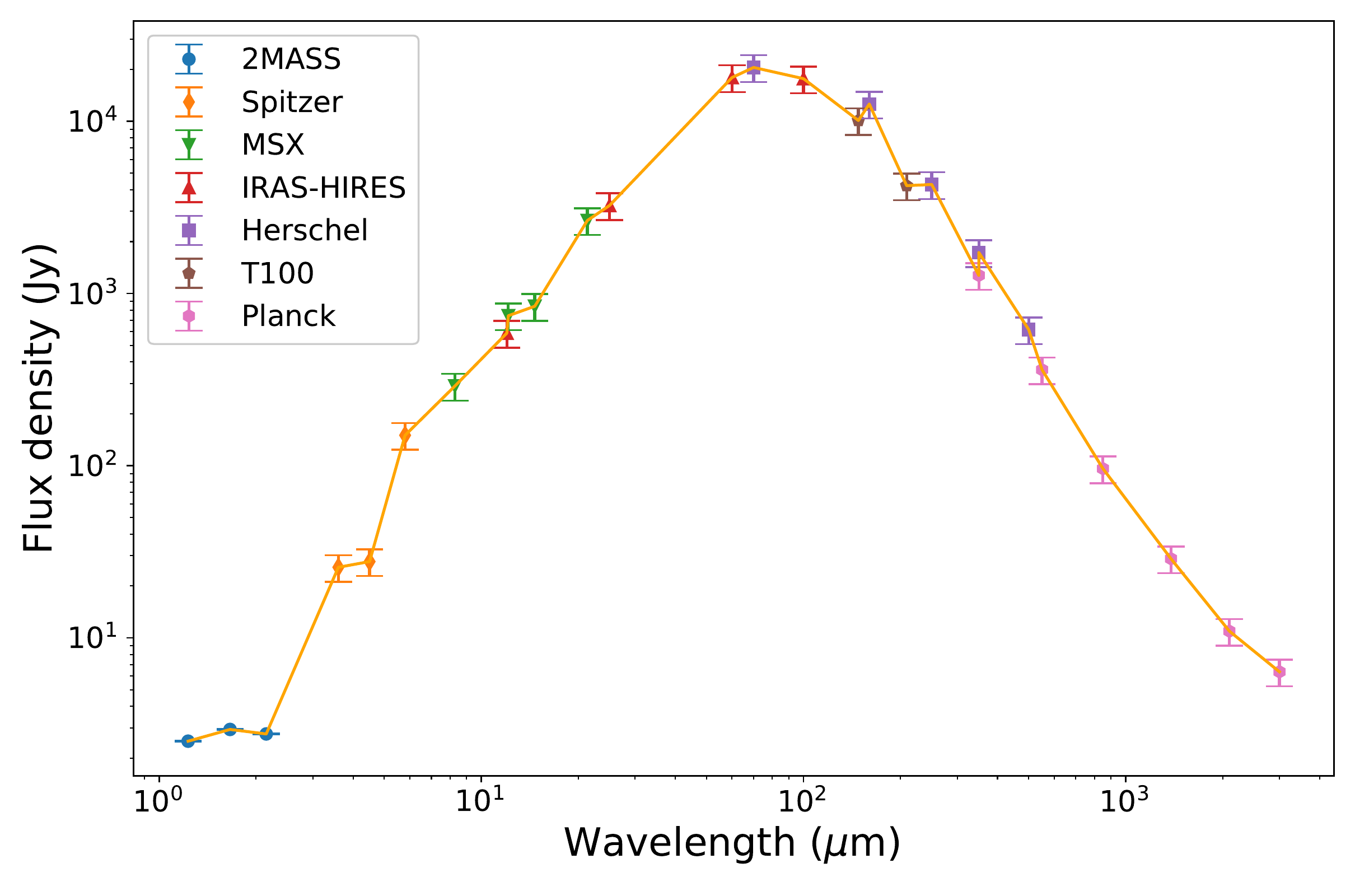}
    \caption{Spectral energy distribution of RCW~42 obtained by integrating emission from a region depicted in Fig.~\ref{RGB_dust}. The error bars correspond to 15$\%$ of the flux density.}
\label{SED}
\end{center}
\end{figure}

We constructed the infrared spectral energy distribution of RCW~42 by integrating flux densities from \textit{Two-Micron All Sky Survey} (2MASS), \textit{Spitzer}-IRAC, IRAS-HIRES \citep{1990AJ.....99.1674A}, \textit{Midcourse Space Experiment} \citep{2001AJ....121.2819P}, \textit{Herschel} Hi-GAL, T100 and \textit{Planck} \citep{2020A&A...641A...1P} images. We integrated emission from a region within an intensity contour corresponding to 5~Jy/pixel of the 70~$\mu$m image, shown in Fig.~\ref{RGB_dust}. The SED of RCW~42 is shown in Fig.~\ref{SED}. We estimate the bolometric luminosity of this region by integrating under the SED obtained by using a power-law interpolation between the flux densities at the observed wavelengths. The estimate of bolometric luminosity of RCW~42 is $1.8\times 10^6$~L$_\odot$. This corresponds to a single ZAMS star of spectral type that is earlier than O4~\citep{1973AJ.....78..929P}, which is consistent with the ZAMS spectral type estimated from ionized gas emission in Sect. 3.1.  

We compare this with the bolometric luminosity of $7.2\ \times\ 10^5$ L$_\odot$ obtained earlier by~\cite{2014MNRAS.437.1791U}, which is comparable to the FIR luminosity from IRAS flux densities \citep{1989AAS...80..149W}. The luminosity value obtained by us is nearly two times larger than this estimate. We note that the luminosity estimate by \cite{2014MNRAS.437.1791U} was obtained by scaling the MSX 21~$\mu$m flux of the region, while the present estimate uses an approach that is more direct while covering a wider range of wavelengths. The FIR luminosity from clumps is similar to the IRAS luminosity estimated earlier. A comparison of the bolometric luminosity with the far-infrared luminosity estimated from the clumps shows that the latter is lower by a factor of nearly 2. 
The differences can be attributed to sampling diffuse regions, and the wavelength coverage. A caveat is that the lower wavelength bands of near-infrared have stellar contribution in the field and this includes stars that do not belong to the RCW~42 complex. However, we anticipate this contribution to be negligible when we compare to the total emission from this region.


\subsection{Ionizing candidates of RCW~42}\label{results_yso}


\begin{figure}
\centering
    \includegraphics[height=7.3cm]{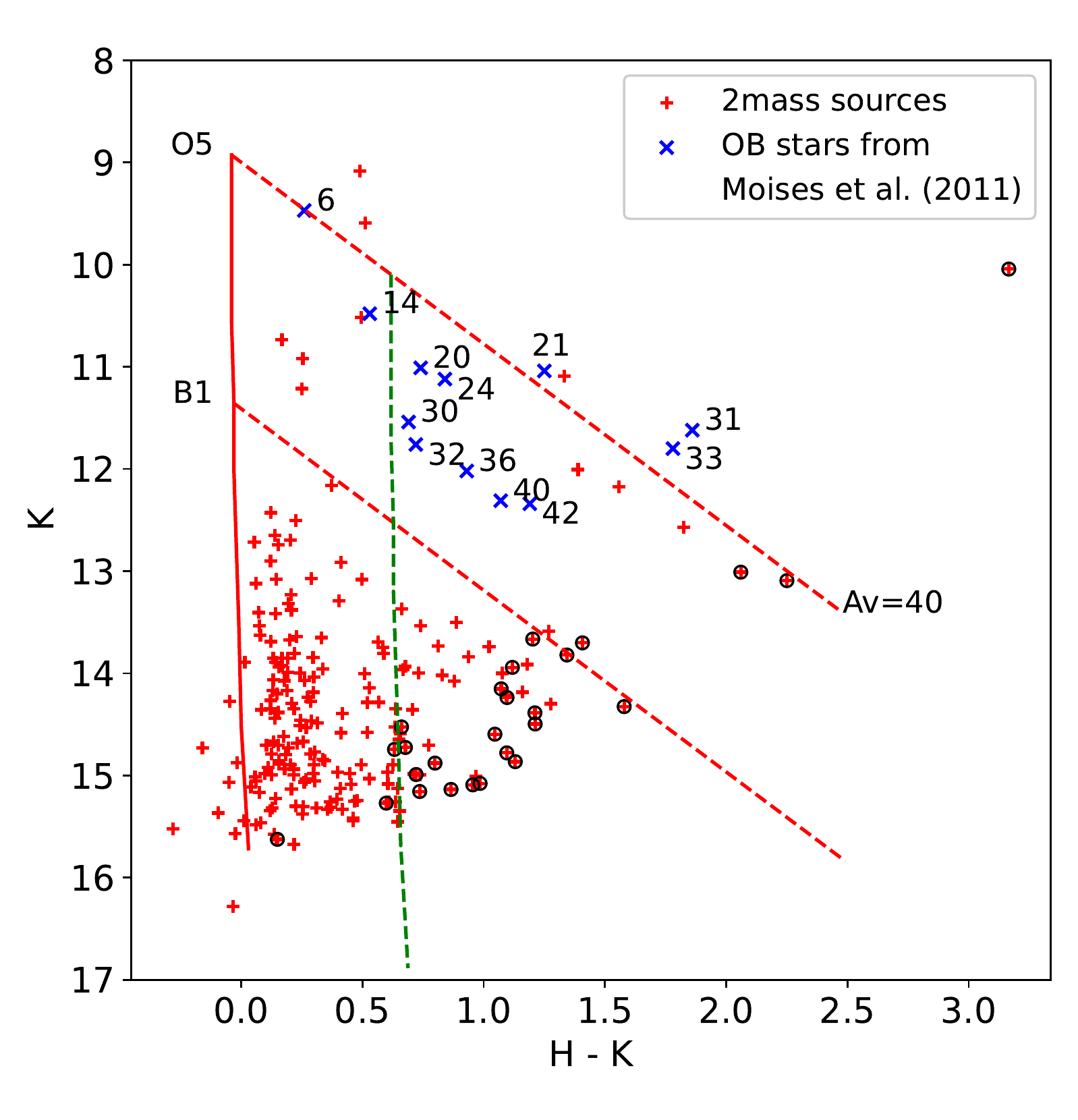}
	\includegraphics[height=7.3cm]{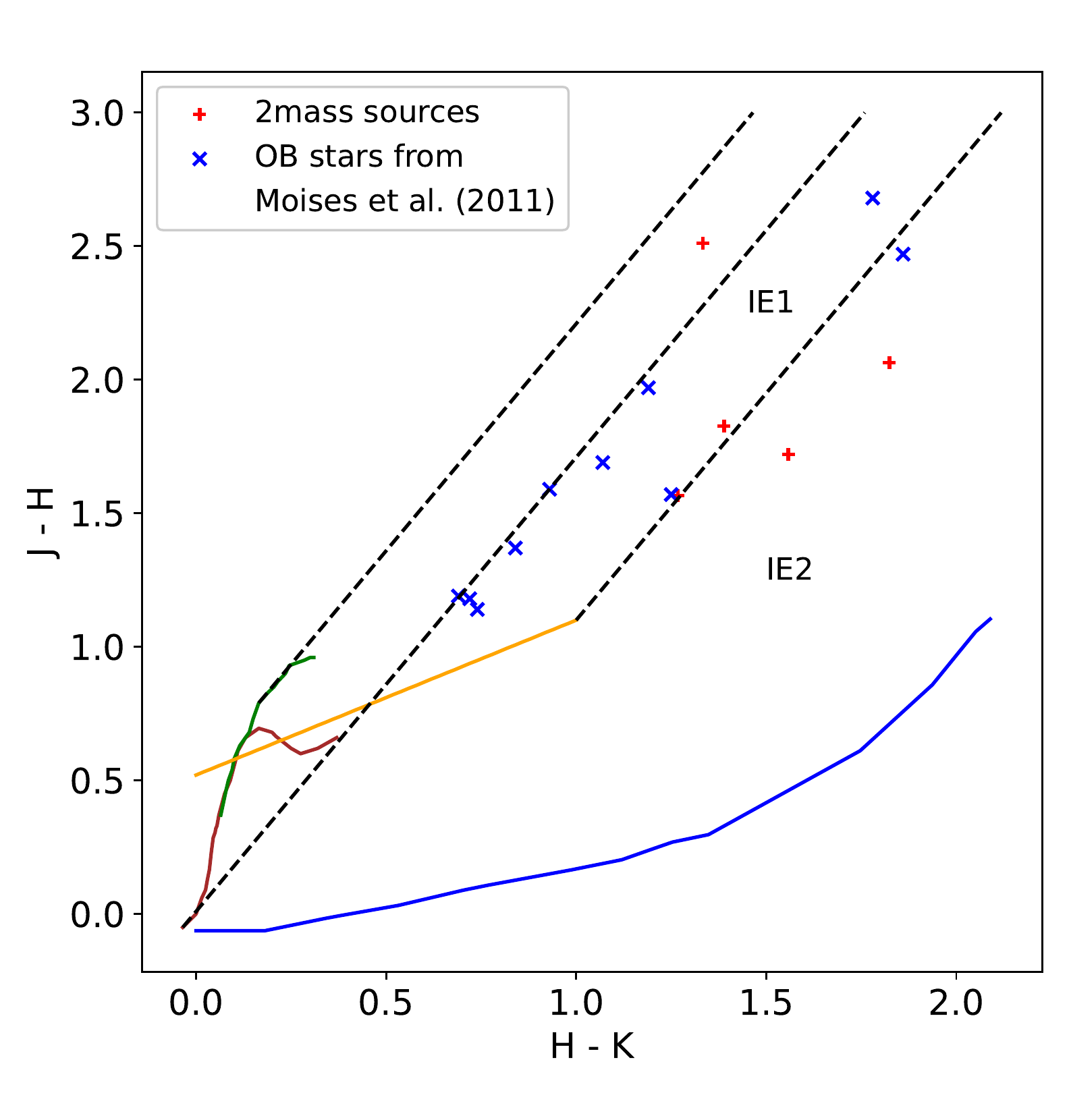}
    \caption{2MASS (a) Color-magnitude diagram (CM-D) and (b) Color-color diagram (CC-D) of point sources within the 5$\sigma$ contour of the 1280 MHz radio image. The stars identified by \protect\cite{2011MNRAS.411..705M} are shown in blue, while the ionizing candidates identified in this study from 2MASS are shown in red. In (a), the encircled sources are those that are detected in HK bands only while the rest of the sources are those detected in all three JHK bands. The red solid line depicts the ZAMS curve. The parallel diagonal dashed lines represent the reddening vectors corresponding to ZAMS O5 and B1 and the dashed green line corresponds to ${\rm A_V}$=10.4 mag. In (b), the loci of the main-sequence stars and giants are shown by the brown and green lines, respectively. The loci of classical T-Tauri stars and Herbig Ae/Be stars, are denoted by the orange and blue curves, respectively. The three parallel dashed lines follow the reddening vectors of giants, main-sequence stars (or dwarfs) and T-Tauri stars. The reddened infrared excess regions are denoted by IE1 and IE2, respectively.}
\label{ccd_cmd}
\end{figure}


The near-infrared images reveal many YSOs present in this HII region, and we are interested in the identification of the potential ionizing sources. The central $1.5\times1.5'$ region associated with near-infrared nebulosity was investigated by \cite{2011MNRAS.411..705M} and they have identified a number of potential ionizing sources, i.e. OB stars using their color-magnitude and color-color diagrams. However, we note that the radio emission occupies a much larger region and we, therefore, employ the 2MASS survey to identify possible OB stars outside the NIR nebulosity.
	
\begin{table*}
\centering
\caption{Details of candidate OB stars towards RCW~42 identified in this work.}
\label{OB_positions}
\begin{tabular}{ccccc} 
\toprule
\toprule
S. No. & 2MASS Designation & RA (J2000) & DEC (J2000) & Detection bands \\
\midrule
1 & 09243936-5200300  & 141.164041 & -52.008343 & JHK             \\
2 & 09243342-5201077  & 141.139281 & -52.018833  & JHK             \\
3 & 09244080-5201484  & 141.170013 & -52.030136  & JHK             \\
4 & 09244307-5202087  & 141.179481 & -52.035763  & JHK             \\
5 & 09241590-5200437  & 141.066271 & -52.012146  & JHK             \\
6 & 09244204-5202031  & 141.175202 & -52.034222  & HK              \\
7 & 09244254-5201506  & 141.177282 & -52.030724  & HK              \\
8 & 09244241-5202102  & 141.176745 & -52.036167  & HK              \\
9 & 09242495-5201507  & 141.103968 & -52.030766  & HK             \\
\bottomrule
\end{tabular}
\end{table*}
	
Using the 2MASS catalog, we consider near-infrared objects within the 5$\sigma$ contour of the 1280 MHz radio image. We select objects with read flags of `222' and `22' implying good photometric quality of the sources. We plot these sources on a color-magnitude diagram (CM-D), K versus H-K to identify the OB type stars in the region. We find a number of such objects. For possible association with RCW~42, we exclude the sources below the extinction level ${\rm{A_V}}$=10.4 mag as they are likely to be foreground objects. This corresponds to extinction due to the diffuse interstellar dust, for a distance to the source  \citep[1.8 mag/kpc;][]{1992dge..book.....W}. We find a total of 9 sources of ZAMS spectral type B1 or earlier from 2MASS. This is in addition to the 10 sources identified by \cite{2011MNRAS.411..705M} as ionizing candidates. Amongst these 9 2MASS sources, 5 are detected in all three 2MASS bands while 4 are detected only in H and K bands. These are shown in Fig.~\ref{ccd_cmd}(a) and their 2MASS designations and coordinates are given in Table~\ref{OB_positions}. The potential ionizing sources identified by \cite{2011MNRAS.411..705M} are also shown in the figure along with the numbers employed by these authors for their identification.

\begin{figure}
	\centering
	\includegraphics[trim={1 1 1 1},clip,height=6.5cm]{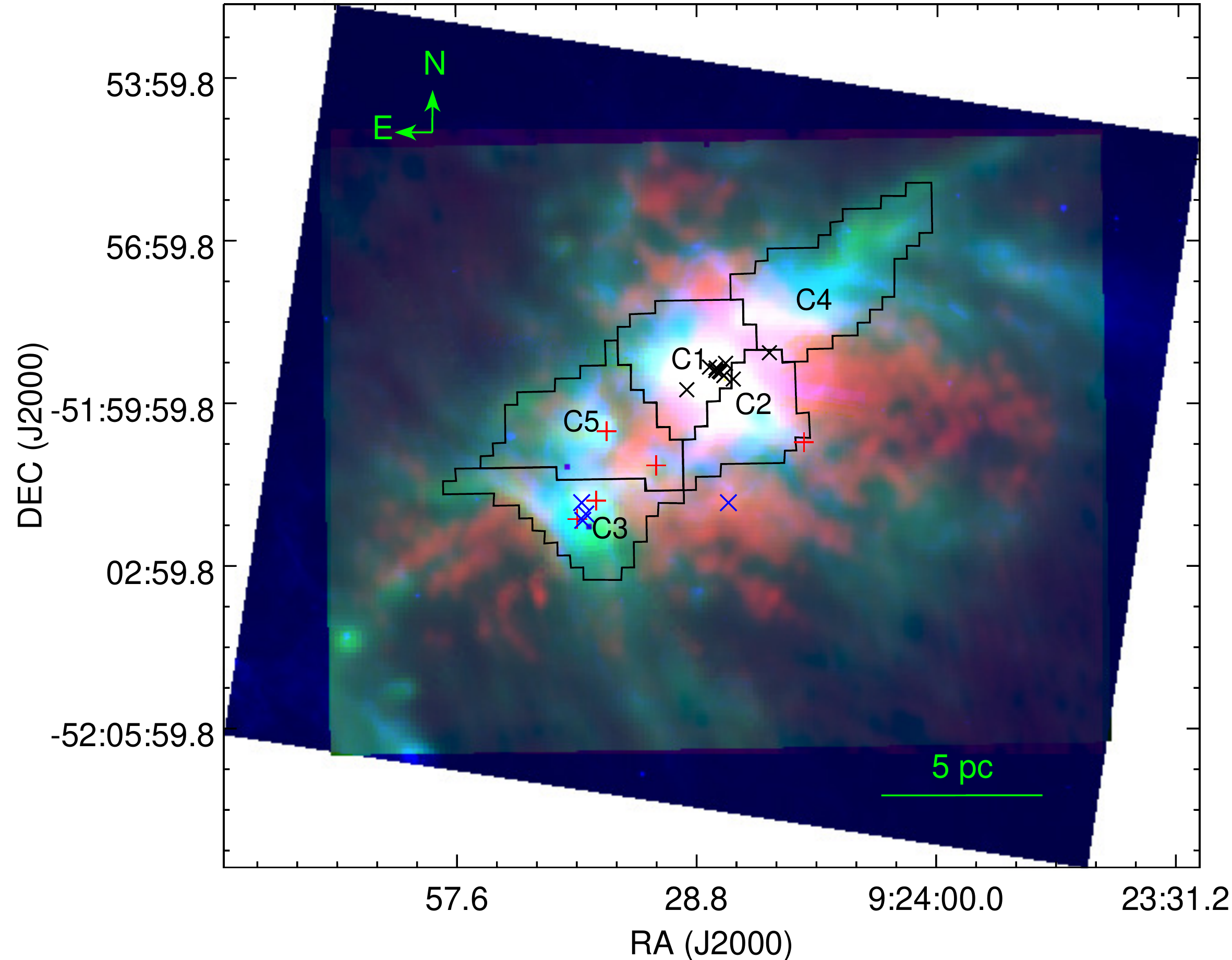}
	\caption{Color-composite image of RCW~42 depicting the ionized gas emission at 1280 MHz (red), cold dust emission at 250~$\mu$m (green) and mid-infrared emission at 8~$\mu$m (blue). The stars identified by \protect\cite{2011MNRAS.411..705M} are shown as black $\times$. The IR excess stars earlier than B1 type and detected in all three 2MASS bands are shown as red + signs. The stars detected in H and K bands and earlier than B1 type are shown as blue $\times$. The clump apertures are also shown.}
	\label{rgb}
\end{figure}

We have plotted the potential ionizing sources associated with RCW~42 in the 2MASS color-color diagram (CC-D) of J-H versus H-K to ascertain the youth of these objects. These are shown in Fig.~\ref{ccd_cmd}(b). The errors associated with sources is $<0.2$~mag. The loci of main-sequence stars, giants, and classical T-Tauri stars are marked \citep{1997AJ....114..288M}. Also shown is the locus of the Herbig Ae/Be stars \citep{1992ApJ...393..278L}. We have used the Bessell and Brett system \citep{1988PASP..100.1134B} for plotting the curves and the magnitudes. We find that barring one source, all the sources lie in the infrared-excess regions depicted as IE1 and IE2 regions in the CC-D. This reinforces the high probability of these candidates being associated with RCW~42. The locations of these sources are shown in Fig.~\ref{rgb}. Of all the sources, only two of them lie outside the clump apertures. It is possible that they are more evolved than the others. There also exists the possibility that they could be background sources. 

A scrutiny of mid-infrared associations of point sources is carried out using the images of the \textit{Spitzer Space Telescope} and the associated catalogs. While the 8~$\mu$m image is saturated, the 5.8~$\mu$m emission shows bright nebulous emission over the entire region. We have inspected the catalogs from the surveys Deep GLIMPSE \citep{2011sptz.prop80074W} and Vela-Carina \citep{2007sptz.prop40791M}. We find a total of 63 mid-infrared sources towards the region occupied by ionized gas emission. The high nebulosity explains the low number of point sources extracted from the catalog towards the radio peak, possibly due to difficulty in locating and segregating the flux densities of the embedded sources in the region. We searched for Spitzer counterparts to the OB candidates identified in the near-infrared that are associated with RCW~42 and we find a total of 6 sources. Two of these are associated with OB candidates identified and mentioned by \cite{2011MNRAS.411..705M} in their work. The remaining four correspond to the OB candidates identified by us in this work.

\section{Discussion}\label{discussion}
In this section, we probe the star formation activity in the HII region following various multiwavelength markers. We discuss the association of the large scale ionized gas with the molecular cloud in this region, and examine the evolutionary stages of the clumps identified in the region.


\subsection{Ionization of gas in the Molecular Cloud}\label{discussion_radioandgas}

The color composite image of the HII region RCW~42 depicting the ionized gas and dust emission is shown in Fig.~\ref{rgb}. One characteristic of the image is that the ionized gas emission is prominent in regions where the dust emission is relatively low. As dust emission is a good proxy for molecular cloud density, it is evident that density gradients have played an important role in the expansion of ionized gas in the cloud \citep{{1978A&A....70..769I},{1979A&A....71...59T}}. From the column density map in Fig.~\ref{dustmaps}, we observe that the densest region of the cloud is elongated with a higher density edge towards the north-east. This is also the direction along which the ionized gas displays a sharp density gradient. We notice that peak positions of all the dust clumps except C4 have mid-infrared and radio counterparts (Figs.~\ref{rgb} and \ref{dustmaps}). This implies the presence of multiple embedded YSOs still embedded in their nascent envelopes. 

A cluster with potential OB star candidates is observed towards RCW~42. If we take the combined ionizing flux of all these OB stars under the assumption that (i) they are ZAMS stars, (ii) NIR colors accurately represent the spectral types, and (iii) all the OB candidates are located at the same distance as RCW~42, we estimate Lyman continuum photon flux rate ${ \dot{N}_{Lyc}}\sim 4 \times 10^{50}$~s$^{-1}$. Although this is about 5 times larger than that estimated from radio continuum emission, considering the simplistic assumptions we have made in order to estimate and compare the values, it would appear that most (if not all) potential OB stars are sources that are likely to ionize RCW~42.

RCW~42 represents a stage in the molecular cloud where the HII region is extended, yet we observe dust clumps. It is, therefore, instructive to obtain estimates of the ionization fraction of the clumps by using the map of electron density $n_e$ described earlier in Sect. 3.1. By taking a ratio of the ionized gas mass determined from the $n_e$ map with the respective total gas mass in each clump aperture, we estimate the ionization fraction of each clump. These ionization fractions are listed in Table~\ref{clumpcharac}. The ionization fraction of clumps lie in the range $2-19\%$. We observe that C1 and C2 have relatively higher ionization fractions of $11$\% and $19\%$, respectively. This is expected for C1 and C2 as the corresponding clump apertures encompass the peak radio intensities. The other clumps, C3, C4 and C5, have lower ionization fractions of $2-5\%$. It is important to take cognisance of the fact that the ionization fractions listed here are averaged over the entire clump and the local values are expected to vary across the clump. If we consider the entire cloud, then the mean ionization fraction is $\sim11\%$.

On scales of tens of parsecs, corresponding to molecular clouds, ionization fraction values are important as they provide hints about the effects of photoionization feedback on star-formation. The expansion of warm photoionized gas and stellar winds from massive stars has long been recognised as a prominent source of feedback within the molecular cloud regulating the formation of star clusters \citep{{2003A&A...411..397T},{2002ApJ...566..302M}}. The expansion of photoionized gas could either compress and fragment the ambient medium facilitating sequential self-propagating star-formation \citep{{2002MNRAS.329..641W},{1977ApJ...214..725E}}, or it could evict gas from the vicinity while unbinding the star cluster \citep{2000ApJ...542..964A}. The average pressure due to the warm gas can be estimated using the ideal gas law, $P=n\mathrm {k_B} T_e$, where $n$ is the average number of particles. The average energy density in the warm gas component is given by the relation $u=2P/3$. The number of particles can be considered as $n=2n_e$ for gas composed of hydrogen and singly ionized helium. Taking $T_e=7900$~K as earlier, we estimate the average energy density across the HII region to be $1.0\times10^{-10}$~erg/cm$^3$ and the total energy of the ionized component is $1.1\times10^{49}$~erg. We believe that this is a very useful estimate for future reference, as it can be used for comparison with other feedback effects in the cloud as described above. 

\begin{figure}
 \centering
  \includegraphics[scale=0.35]{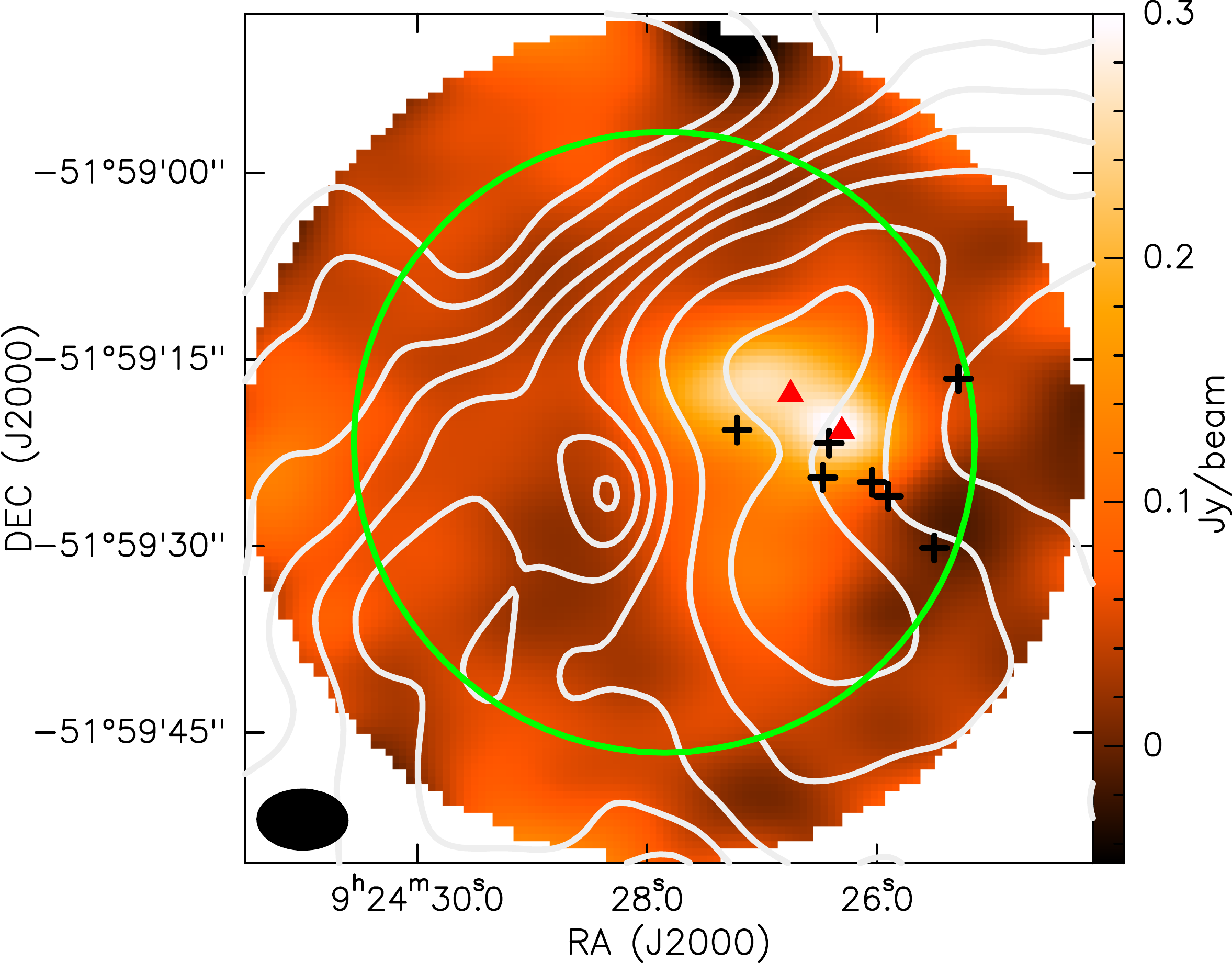}
  \caption{Continuum map of C1 at 1.4 mm overlaid with contours of emission at 1280~MHz. The two radio sources seen in the continuum maps of \protect\cite{1998MNRAS.301..640W} are marked as red triangles. The stars identified by \protect\cite{2011MNRAS.411..705M} that lie within the primary beam (field-of-view) are marked as + signs. The primary beam is marked with a green circle and the synthesized beam (FWHM) is marked at the bottom left. The contour levels are 3.0, 7.2, 15.8, 27.9, 40.8, 58.7, 85, 140, 215 mJy/beam.}
\label{ALMAradio2mass}
\end{figure}


\subsection{Evolutionary Stages of Clumps}\label{discussion_clump_evolution}
We determine the evolutionary stages of the dust clumps in the region by using the evolutionary sequence given by \cite{2010ApJ...721..222B}. The first stage in this sequence involves a quiescent clump that is devoid of any signs of active star formation. This quiescent clump then progresses into an active clump and later turns to an evolved clump. According to the classification, there are four essential star formation traces: (1) Absence of any sign of active star formation in Quiescent clumps, (2) One or two signs of active star formation (a 24 $\mu$m point source or shock/outflow signatures) exhibited by Intermediate clumps, (3) Three or four signs of active star formation (UC HII region, 24 $\mu$m point source, maser emission) exhibited by Active clumps, and (4) Diffuse 8 $\mu$m emission exhibited by Evolved clumps.

As there is no MIPS 24 $\mu$m data available for this region, we employed the WISE 22 $\mu$m data for the clumps. We searched for 22 $\mu$m point sources within a radius of 25$''$ from the peak position of the clump. We find that all the clumps except C1 have 22 $\mu$m point sources close to clump peak positions. The emission towards C1 is saturated. We also searched for masers associated with this region in literature and found that although \cite{1998MNRAS.301..640W} and \cite{1995MNRAS.274..808C} did search for 6.67 GHz methanol maser and 4.7 GHz OH maser, respectively, towards this region, no maser emission was detected. Hence, the classification was  carried out based on radio, 8 $\mu$m and 22 $\mu$m emission as per the scheme of \cite{2010ApJ...721..222B}.

We find that all the clumps are active or evolved with the presence of multiple signs of star formation activity. Radio emission is detected towards all of the clumps which implies that high mass star formation is occurring within these clumps. This is again corroborated from the near-infrared color-magnitude and color-color diagram. All clumps except clump~C4 are associated with sources showing infrared excess whose ZAMS spectral type is earlier than B1. 

Clump~C1 is the most active amongst the clumps. It hosts a cluster of massive stars, as seen from Fig.~\ref{ccd_cmd}. There are 8 stars of ZAMS spectral type earlier than B1 clustered within 0.6~pc towards the radio emission peak. The peak of radio emission appears elongated along N-S direction. The 1.4~mm continuum emission also depicts similar elongation. This can be seen in Fig.~\ref{ALMAradio2mass}. The cores MM1 and MM2 lie close to the northern lobe of ionized gas emission. High resolution radio images \citep{1998MNRAS.301..640W} imply that both these cores are associated with UCHII regions. C1 also harbours the IRAS source I-09227. Clump~C3 is also associated with an IRAS source I-09230. It has radio emission with elongated morphology that is associated with extended green object EGO G274.0649--01.1460, suggesting the presence of high mass protostellar outflow(s) within C3. The morphology of the radio emission could be a direct evidence of ionized jet from the central source. Clumps C2, C3, C4 and C5 harbour multiple 22~$\mu$m sources strongly indicating clustered star formation within these clumps, similar to that of C1.


\begin{figure}
	\centering
	\includegraphics[trim={1 1 1 1},clip,scale=.4]{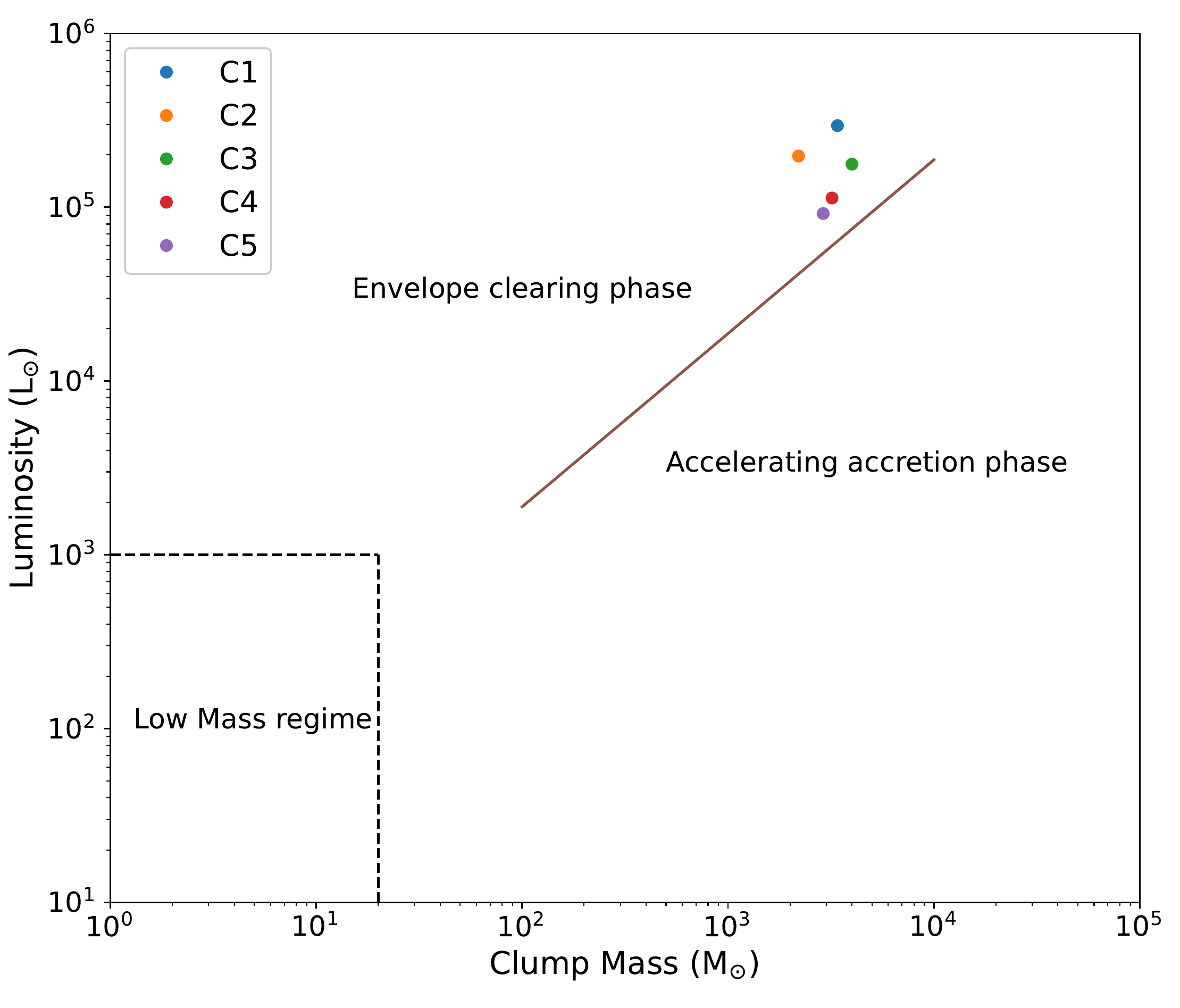}
	\caption{Mass-luminosity plot for the clumps. The locus that distinguishes the final-envelope clearing phase from the accelerating accretion phase according to the models of \protect\cite{2008A&A...481..345M} is shown as a solid line. The identification of the low-mass regime is based on \protect\cite{1996A&A...309..827S}.}
	\label{mass_luminosity}
\end{figure}

The evolutionary stages of the clumps can also be visualized using a mass-luminosity (M-L) diagram. This diagram was initially proposed by \cite{1996A&A...309..827S} for the evolutionary stages of star formation in the low mass regime and later modified by \cite{2008A&A...481..345M} for the high mass regime as well. The evolution of a massive clump begins in the lower right part of the diagram with high mass and low luminosity. As the source evolves, it moves almost vertically up in the diagram with an increase in luminosity but no significant change in the mass. This is a result of the increasing protostellar activity with evolution. The source is said to be in the accelerating accretion phase in this stage. Once the source reaches the ZAMS stage, it starts moving to the left in an almost horizontal manner with a decrease in mass and no significant change in luminosity. This is the envelope clearing phase. The mass-luminosity plot for the five clumps identified in this HII region is shown in Fig.~\ref{mass_luminosity}. The solid line shown in the plot marks the separation between the accelerating accretion phase and the envelope clearing phase \cite[]{2008A&A...481..345M}. From the figure, it is evident that all of the clumps in RCW~42 fall in the high mass regime. In addition, the positions of these clumps in the M-L diagram indicate that all these clumps are in the envelope clearing phase. This result is in well agreement with the classification of clumps based on star formation tracers. Overall, RCW~42 represents a star-forming region where all the clumps are actively forming massive stars.

\section{Conclusions}\label{conclusion}
We arrive at the following conclusions in our study of RCW~42.
\begin{enumerate}
\item The radio emission from the HII region associated with RCW~42 shows large scale emission extending upto $20\times15$~pc$^2$. The emission shows a steep density gradient towards the eastern side. We have constructed the electron density map and we estimate the average ionization fraction of the cloud to be $11\%$.
\item We construct column density and temperature maps of the region using FIR data from the Herschel archive. The peak column density and temperature values for the region are 5.83 $\times$ 10$^{22}$ cm$^{-2}$ and 39.1 K.
\item Five high density dust clumps are identified in the region. The clump masses range from 2200 M$_{\odot}$ to 4000 M$_{\odot}$ whereas the clump luminosities range from 90 - 300 $\times$ 10$^3$ L$_{\odot}$. The total cloud mass corresponds to $1.6 \times 10^4$~M$_{\odot}$ and the total cloud luminosity is 1.8 $\times$ 10$^6$ L$_{\odot}$.
\item We have identified an extended green object EGO G274.0649--01.1460 as well as several YSOs in the region within the clumps.
\item All the 5 dust clumps identified in the region are either in active or evolved phase. This implies that RCW~42 is in a relatively advanced evolutionary phase.

\end{enumerate}

\section*{Acknowledgements}\label{acknowledgement}
\small
We thank the staff of the GMRT that made these observations possible. GMRT is run by the National Centre for Radio Astrophysics of the Tata Institute of Fundamental Research. We thank the members of the Far Infrared Astronomy programme at TIFR who developed the Infrared telescope T100 and the members of TIFR Balloon Facility, Hyderabad for their roles in conducting the balloon flights. SV acknowledges support from the Dept. of Science and Technology - SERB grant CRG/2019/002581. VVS acknowledges support from the Alexander von Humboldt Foundation. DKO acknowledges the support of the Department of Atomic Energy, Government of India, under Project Identification No. RTI 4002.

This publication makes use of data products from the Two Micron All Sky Survey, which is a joint project of the University of Massachusetts and the Infrared Processing and Analysis Center/California Institute of Technology, funded by the National Aeronautics and Space Administration and the National Science Foundation. This work is based [in part] on observations made with the Spitzer Space Telescope, which was operated by the Jet Propulsion Laboratory, California Institute of Technology under a contract with NASA. This publication also made use of data products from \textit{Herschel}. \textit{Herschel} is an ESA space observatory with science instruments provided by European-led Principal Investigator consortia and with important participation from NASA. This paper makes use of the following ALMA data: ADS/JAO.ALMA$\#$2019.1.00195.L. ALMA is a partnership of ESO (representing its member states), NSF (USA) and NINS (Japan), together with NRC (Canada), MOST and ASIAA (Taiwan), and KASI (Republic of Korea), in cooperation with the Republic of Chile. The Joint ALMA Observatory is operated by ESO, AUI/NRAO and NAOJ.

\section*{Data availability}
\small
The original data underlying this article will be shared on reasonable request to the corresponding author.



\bibliographystyle{mnras}
\bibliography{rcw42_ref} 


\bsp	
\label{lastpage}
\end{document}